\documentclass[prd,twocolumn,showpacs]{revtex4}
\usepackage{bm}
\usepackage{graphicx}
\begin{document}

\newcommand{\mwimp}{$m_\chi$}
\newcommand{\sigmapsi}{$\sigma_p^{SI}$}
\newcommand{\sigmapsd}{$\sigma_p^{SD}$}
\newcommand{\sigmansd}{$\sigma_n^{SD}$}
\newcommand{\kms}{km~s$^{-1}$}
\newcommand{\vsun}{$v_\odot$}
\newcommand{\ve}{$v_\oplus$}
\newcommand{\vlag}{$v_{\mathrm{lag}}$}
\newcommand{\vrms}{$v_{\mathrm{rms}}$}
\newcommand{\vescsun}{$v^{\odot}_{esc}$}
\newcommand{\vesce}{$v^{\oplus}_{esc}$}
\newcommand{\Qmax}{$Q_{\mathrm{max}}$}
\newcommand{\Qmin}{$Q_{\mathrm{min}}$}

\title{WIMP astronomy and particle physics with liquid-noble and cryogenic direct-detection experiments}
\author{Annika H.\ G.\ Peter}
\email{annika.peter@uci.edu}
\affiliation{Department of Physics and Astronomy, University of California, Irvine, California  92697-4575, USA\\California Institute of Technology, Mail Code 249-17, Pasadena, California 91125, USA}

\begin{abstract}
Once weakly-interacting massive particles (WIMPs) are unambiguously detected in direct-detection experiments, the challenge will be to determine what one may infer from the data.  Here, I examine the prospects for reconstructing the local speed distribution of WIMPs in addition to WIMP particle-physics properties (mass, cross sections) from next-generation cryogenic and liquid-noble direct-detection experiments.  I find that the common method of fixing the form of the velocity distribution when estimating constraints on WIMP mass and cross sections means losing out on the information on the speed distribution contained in the data and may lead to biases in the inferred values of the particle-physics parameters.  I show that using a more general, empirical form of the speed distribution can lead to good constraints on the speed distribution.  Moreover, one can use Bayesian model-selection criteria to determine if a theoretically-inspired functional form for the speed distribution (such as a Maxwell-Boltzmann distribution) fits better than an empirical model.  The shape of the degeneracy between WIMP mass and cross sections and their offset from the true values of those parameters depends on the hypothesis for the speed distribution, which has significant implications for consistency checks between direct-detection and collider data.  In addition, I find that the uncertainties on theoretical parameters depends sensitively on the upper end of the energy range used for WIMP searches.  Better constraints on the WIMP particle-physics parameters and speed distribution are obtained if the WIMP search is extended to higher energy ($\sim 1\hbox{ MeV}$).
\end{abstract}


\maketitle

\section{Introduction}\label{sec:intro}
Dark matter makes up $\sim 23\%$ of the energy density of the observable Universe, yet its identity is unknown (e.g., \cite{komatsu2010}).  While there are a number of well-motivated particle-physics candidates for dark matter (e.g., \cite{turner1990,dodelson1994,abazajian2001,feng2004,feng2008,arkanihamed2009,kaplan2009}), the most popular particle class is the weakly-interacting massive particle (WIMP) \cite{steigman1985}.  This class of dark-matter candidate is popular because a number of particles in this class arise ``for free'' and at the right relic abundance in extensions to the standard model \cite{griest1988b}.  Moreover, due to their weak but non-negligible coupling to standard-model particles, it is possible to detect them.  Candidates in this class include the supersymmetric neutralino and the Kaluza-Klein photon \cite{griest1988b,jungman1996,servant2002,cheng2002}.  

There is a wide variety of efforts focused on finding and characterizing WIMP dark matter, which can be broadly classified as creating (in colliders), destroying (by annihilation in dark-matter-dense astrophysical objects), or colliding with WIMPs (using nuclei in terrestrial detectors) \cite{goodman1985,krauss1985,silk1985,wasserman1986,baer1987,griest1988b,bengtsson1990,mrenna1996}.  This last method, called ``direct detection'', is the subject of this work.  There is a broad ongoing effort to find and identify WIMPs using direct-detection experiments.  Currently, only the DAMA/LIBRA collaboration claims a direct detection of dark matter \cite{bernabei2010}, a controversial claim given the nondetections from other experiments \cite{cdms2010,cdmslowe2010,aprile2010}.  Experimental efforts can be roughly divided between those focused on detecting WIMPs through their spin-dependent (axial-vector) couplings to nuclei and those focusing on spin-independent scattering on nuclei.  

The most mature technologies are those associated with searches for spin-independent (SI) WIMP-nucleon scatters.  Cryogenic experiments such as CDMS, Edelweiss II, CRESST, and CoGeNT can distinguish nuclear from electronic recoils using different (ionization, scintillation, and heat) signals \cite{lee2007,aalseth2008,aalseth2009,edelweiss2010,mckinsey2010,eureca2010}.  Liquid-noble gas experiments such as XENON100, LUX, XMASS, WArP, ArDM, DEAP/CLEAN, DarkSide, and Zeplin-III can distinguish between the two types of recoils using a combination of the amount of scintillation light, ionization yield, pulse shape, and timing \cite{ardm2010,giuliani2010,minamino2010,warp2010,baudis2010,zeplin2010}.  These experiments can resolve the energies but not directions of the recoils.  The current best limits on the spin-independent WIMP-proton cross section (\sigmapsi) arise from using $\lesssim 1000 \hbox{ kg}\cdot\hbox{day}$ of data, and at are the level of $\sigma_P^{SI} \lesssim 4\times 10^{-44}\hbox{ cm}^2$ for a WIMP mass \mwimp$\approx 50$ GeV.  The targets for these experiments are increasing rapidly, with $\sim$ton-scale liquid-noble and $\sim100$ kg cryogenic experiments expected to be operational within the next five years (in or around year 2015) \cite{baudis2010,giuliani2010,bruch2010}.  Experiments an order of magnitude bigger than those are being discussed, to be constructed approximately ten years from now \cite{baudis2010,eureca2010,golwala2009,gaitskell2010}.  Those 2020- to 2025-era experiments should have WIMP sensitivities 4 to 5 orders of magnitude better than those today.  

The question is, if these next-generation direct-detection experiments see unambiguous WIMP signals, what will we learn about WIMPs from them?  Most of the effort thus far has been focused on determining how well one may infer the WIMP mass and cross sections.  These are fundamental particle-physics WIMP parameters that will allow us, in combination with indirect detection and production at colliders, to determine to which extension to the standard model the WIMP belongs.  However, the energy spectrum of events in direct-detection experiments depends not only on the WIMP mass and cross sections, but also on the dark-matter distribution function (DF).  Thus, any inference of the WIMP mass and cross sections from the data also depends on the DF (see Eq. \ref{eq:drdq} in Sec. \ref{sec:methods}).  The WIMP DF is typically modeled with a fixed theoretically-inspired form (e.g., an isotropic Maxwell-Boltzmann distribution or direct fits to N-body simulations) in which the parameters of the model (e.g., the one-dimensional velocity dispersion \vrms) are either fixed or only allowed to vary in a narrow range \cite{green2008a,strigari2009,alves2010,peter2010b,green2010,pato2010,akrami2010b,lisanti2011}.  Implicit in this treatment of the WIMP DF is that it is well described by a globally smooth dark-matter halo model.

However, the actual local dark-matter DF is unknown.  Even if the local dark-matter DF is dominated by a halo component, we do not know exactly what to expect.  High-resolution dark-matter-only N-body simulations indicate significant halo-to-halo variation in the DF of the smooth component of the halo as well as $\sim\hbox{ kpc}$ scale fluctuations in the DF that are dynamically cold imprints of the halo accretion history \cite{vogelsberger2009,kuhlen2010}.  The velocity distribution is typically anisotropic.  In addition to a smooth halo component to the local DF, there could also be significant contributions from a dark disk \cite{read2008,bruch2009a,read2009,purcell2009,ling2010} or small-scale velocity streams (below the resolution limit of simulations) that have not yet phase mixed.  

Direct-detection experiments and neutrino searches for WIMP annihilation in the Sun and Earth are the \emph{only} probes of the local DF of WIMPs, unless there is a significant velocity-dependence in the annihilation cross section.  While there have recently been some attempts to constrain the WIMP mass and cross sections by ``integrating out'' the uncertainty in the WIMP velocity distribution \cite{fox2010,fox2011}, it is highly desirable to use the direct-detection data to understand the WIMP DF as well as the particle-physics properties of dark matter.

In this work, I explore the prospects for determining the WIMP speed distribution (the integral of the DF over configuration-space volume and velocity orientation) for several benchmark points in \mwimp$-$\sigmapsi space and velocity distribution models from 2015-era cryogenic and liquid-noble direct-detection experiments.  Using a Bayesian framework to analyze mock data sets, I show that one may infer the WIMP speed distribution as well as the WIMP mass and cross section from even a modest number of events, assuming that WIMP events are identified in at least one 2015-era direct-detection experiment.  

I consider several scenarios.  First, I show how well one may characterize the speed distribution as well as the WIMP particle-physics parameters if the hypothesis for the speed distribution matches the data, but for which the parameter values of the hypothesis are previously unknown.  Since parameter constraints are most accurate and unbiased if the hypothesis is correct, this is a demonstration of the best constraints we can get from the data.  Second, I consider the case in which the hypothesis for the speed distribution is wrong, as would be the case if the local WIMP population had dark-disk and stream components in addition to a smooth halo component, but one were to analyze data with the hypothesis that only a single velocity component exists.  Finally, I show the constraints one obtains on the WIMP mass, cross sections, and speed distribution with the hypothesis of a simple empirical form for the speed distribution.  This is a proof of principle of the usefulness of empirical speed-distribution models for parameter estimation and Bayesian model selection. While this is not the first exploration of empirical treatments of the WIMP speed distribution \cite{drees2007,drees2008,shan2011a,shan2011b}, the unbinned likelihood and the Bayesian framework I employ below have the advantage of being easily modified to incorporate backgrounds, systematic errors, and additional data sets of various types (not limited to direct detection).  In addition, this work highlights the importance of the hypothesis for the form of the WIMP speed distribution in inferring WIMP particle-physics parameters from direct-detection data.

The outline of this paper is as follows.  In Sec. \ref{sec:methods}, I describe the ans{\"a}tze and methods used to infer WIMP properties and the speed distribution from mock data sets.  In Sec. \ref{sec:results}, I apply the methods in Sec. \ref{sec:methods} to mock data sets for a set of WIMP particle-physics and speed-distribution benchmarks.  In Sec. \ref{sec:discussion}, I discuss the implications of the results of Sec. \ref{sec:results} for estimating the local WIMP speed distribution in the future, and discuss the results in the context of WIMP characterization using a combination of data sets, including those from the Large Hadron Collider.  The key points of this work are summarized in Sec. \ref{sec:conclusion}.

\section{Ansatz \& Method}\label{sec:methods}
The plan is to estimate how well one may reconstruct the WIMP speed distribution as well as the particle-physics properties of WIMPs (mass, cross sections) in 2015-era liquid-noble and cryogenic direct-detection experiments.  These experiments can resolve the energy of WIMP-induced nuclear recoils but not the direction of the recoiling nucleus.

In the absence of energy errors, the differential event rate per kilogram of a target $N$ with nuclear mass $m_N$ in an direct-detection experiment is
\begin{eqnarray}\label{eq:drdq}
  \frac{dR}{dQ} = \left( \frac{m_N}{\mathrm{kg}} \right)^{-1}  \int_{v_{\mathrm{min}}} d^3 \mathbf{v} \frac{d \sigma_N}{d Q} v f(\mathbf{x},\mathbf{v}),
\end{eqnarray}
where $d\sigma_N/dQ$ is the differential scattering cross section, $f(\mathbf{x},\mathbf{v})$ is the local dark-matter DF, and
\begin{eqnarray}
v_{\mathrm{min}} = ( m_N Q / 2 \mu^2_N)^{1/2} \label{eq:vmin}	
\end{eqnarray}
is the minimum speed required for a particle of mass $m_\chi$ to deposit energy $Q$ to the nucleus if the interaction is elastic, and $\mu_N$ is the WIMP-nucleus reduced mass.  In this work, I assume that the interactions are elastic, deferring the discussion of inelastic interactions to future work \cite{peter2011b}.  

If one neglects the annual modulation of the direct-detection signal (due to the Earth's motion relative to the WIMP velocity distribution), the integral over the WIMP directions is time independent, and one can thus consider the \emph{speed} distribution of WIMPs rather than the full velocity distribution.  The speed distribution $g(v)$ is defined such that
\begin{eqnarray}\label{eq:gv}
  g(v) &=& \int d\Omega_v f(\mathbf{x},\mathbf{v}) / n_\chi\\
  \int g(v) v^2 dv &=& 1,\label{eq:gvlim}
\end{eqnarray} 
where $n_\chi = \rho_\chi / m_\chi$ is the number density of dark-matter particles.  Implicit in Eq. (\ref{eq:gv}) is the assumption that the number density is constant throughout the duration of the experiment, i.e., that the number density does not vary significantly along the Earth's path through the Solar System.  Annual modulation provides an interesting constraint on the full velocity distribution, not just the speed distribution, but I will defer a discussion of this to future work.

I create mock direct-detection data sets using a variety of particle-physics (Sec. \ref{subsec:particle}) and speed-distribution benchmarks (Sec. \ref{subsec:astro}) for a set of toy-model 2015-era experiments (Sec. \ref{subsec:toy}).  I estimate particle-physics and speed-distribution parameters from the mock data sets using the likelihood and sampling techniques described in Sec. \ref{subsec:parameter}.

\subsection{Particle physics}\label{subsec:particle}

For the time being, I assume that the spin-dependent (SD) WIMP-proton cross section \sigmapsd$= 0$, and that all events result from spin-independent elastic scattering.  The scattering cross section for Eq. (\ref{eq:drdq}) for a target nucleus with atomic number $A$ is thus (e.g., Ref. \cite{jungman1996})
\begin{eqnarray}
  \frac{d\sigma_A}{dQ} = \frac{m_A}{2 v^2 \mu_p^2} A^2 \sigma_p^{SI} F_{SI}^2(Q),
\end{eqnarray}
where $m_A$ is the nuclear mass, $\mu_p$ is the reduced mass of the WIMP-proton system, and $F_{SI}(Q)$ is the nuclear form factor.  I assume that the coupling of WIMPs to protons is identical to the WIMP-neutron coupling, and use a Helm form factor for $F_{SI}$ \cite{helm1956}.

\subsection{Astrophysics}\label{subsec:astro}
As benchmark models for the mock experiments, I take one or more isotropic Maxwell-Boltzmann distributions, which in a frame corotating with the Earth have the form
\begin{eqnarray}
  f(\mathbf{v}) = \frac{\rho_\chi/m_\chi}{(2\pi v_\mathrm{rms}^2)^{3/2}} e^{-(\mathbf{v}-\mathbf{v_{\mathrm{lag}}})^2/2v_{\mathrm{rms}}^2}.\label{eq:MB}
\end{eqnarray}
Here, $\rho_\chi$ is the local WIMP density, \vrms~is the one-dimensional velocity dispersion of particles, and \vlag~is the relative speed of the center of the Maxwell-Boltzmann distribution with respect to the experiments.  The astrophysical reason for choosing this model for the WIMP velocities is described in Sec. \ref{subsec:1mb}.  I choose to use distributions that are isotropic in the rest frame of the WIMPs for simplicity, although, in general, anisotropic velocity distributions are expected \cite{vogelsberger2009}.  In principle, one can input an arbitrary speed distribution to an analysis of the type done in Sec. \ref{sec:results}, but that is beyond the scope of this work.

For this work, I do not cut off the DF above an escape velocity $v_{\mathrm{esc}}$ from the Galaxy, although this is an easy thing to add.  The key points of this work hold regardless of the inclusion or exclusion of $v_{\mathrm{esc}}$ in the DF.  Moreover, there may be WIMPs passing through the experiments that lie above the escape speed, as the Milky Way is certainly not in dynamical equilibrium \cite{freese2001,belokurov2007,bell2008}.  I also neglect the effect of gravitational focusing due to the gravitational potential wells of the Earth and Sun.  However, gravitational focusing is most relevant for WIMPs with speeds $v \lesssim 100\hbox{ km s}^{-1}$, which, as I show in Sec. \ref{subsec:vel}, are not generally accessible to the types of experiments described in Sec. \ref{subsec:toy}.

I define the ``standard halo model'' (SHM) as a single Maxwell-Boltzmann distribution with \vlag$=220\hbox{ km s}^{-1}$, which is the IAU value for the speed of the local standard of rest (LSR) \cite{kerr1986}.  This value is $\sim 10\%$ lower than that inferred from recent astrometric measurements of masers in star-forming regions in the Milky Way \cite{reid2009,bovy2009,mcmillan2010}.  The rms speed for the SHM is taken to be \vrms=\vlag$/\sqrt{2}$.  The factor of $\sqrt{2}$ arises from the Jeans equation if one approximates the density profile of the galactic halo as $\rho(r) \propto r^{-2}$ and the rotation curve as flat (see Ref. \cite{binney2008}, Appendix A of Ref. \cite{peter2008}, and Sec. \ref{subsec:1mb}).

Simulations of disk galaxies in dark-matter halos show that massive satellites are preferentially dragged into the disk plane, where they subsequently disrupt due to tidal forces \cite{lake1989,read2008,read2009,purcell2009,ling2010}.  The disrupted dark matter settles into flared-disk-like structure coincident on the baryonic disk, thus forming a ``dark disk.''  Disk galaxies are generically expected to have dark disks, although the properties of the disk depend strongly on the accretion history of the host halo.  Thus, we expect that the local WIMP DF should have a dark-disk component.  Using Ref. \cite{read2009} as a guide, I define a ``standard dark disk'' (SDD) velocity distribution as having the form of Eq. (\ref{eq:MB}) with \vlag$=100\hbox{ km s}^{-1}$ and \vrms$=50\hbox{ km s}^{-1}$.  The weight of the SDD with respect to the halo models will be described in Sec. \ref{subsec:2mb} in which I consider multimodal speed distributions.

\subsection{Toy experiments}\label{subsec:toy}
I simulate data sets for four idealized 2015-era experiments.  The first two experiments use liquid xenon as their target material, inspired by the planned XENON1T and under-construction LUX experiments \cite{aprile2009b,mckinsey2010}.  The third toy experiment is based on the several ton-scale liquid-argon experiments planned and under construction (e.g., the various experiments in the DEAP/CLEAN program \cite{giuliani2010}, ArDM \cite{ardm2010}).  The last toy experiment is based on the under-construction SuperCDMS cryogenic germanium experiment \cite{bruch2010}.  I assume a total xenon mass of 1 ton and an exposure of 1 year for the ``XENON1T-like'' experiment, 350 kg of xenon and an exposure of 1 year for the ``LUX-like'' experiment, 1 ton of argon and an exposure time of 1 year for the ``argon'' experiment, and 100 kg of germanium and a 1-year exposure for the ``SuperCDMS-like'' experiment.

Note that I do not consider constraints on the speed distribution and WIMP particle-physics parameters for each experiment individually.  As I \cite{peter2010b} and others \cite{drees2008,pato2010} have shown, unless one fixes either the WIMP mass or the speed distribution of WIMPs \emph{a priori}, one does not obtain meaningful parameter estimates from a single experiment.  This is because one needs to have a handle on what sets the energy scale for the recoils: the WIMP mass or the WIMP speeds, since the recoil energy is given by
\begin{eqnarray}
  Q = \frac{\mu_A^2}{m_A}v^2 (1 - \cos\theta),
\end{eqnarray}
where $\theta$ is the center-of-mass scattering angle.  The true power comes in having a variety of experiments with different target nuclei, which allows one to break the degeneracy between WIMP mass and WIMP speeds in the recoil energy spectrum.  Moreover, many experiments do and will continue to run simultaneously, and there is no reason not to consider the combined constraints from all experiments.

These toy experiments are idealized in that I assume that backgrounds are negligible, and that they have perfect energy resolution and no systematic errors.  The reasons for choosing such idealized scenarios are the following.  First, the actual background rates and energy resolution for the 2015-era experiments are unknown, although the goal of most experiments is to get to the zero-background regime.  Energy errors for the current germanium-based experiments are negligible for parameter-estimation purposes \cite{peter2010b}, but are potentially a major issue for liquid-noble experiments.  For example, there is currently a large systematic error on the inferred nuclear recoil energies based on the scintillation light observed in xenon-based experiments \cite{aprile2009,aprile2010,manalaysay2010,manzur2010,sorensen2011}.  Experiments are underway to better characterize the relation between the energy seen in experiments and the nuclear recoil spectrum, so it is likely that the energy resolution, systematics, and background sources will be far better characterized in the future than they are now.  Second, by using idealized experiments, I show the minimum expected uncertainty in the WIMP parameters.  Any backgrounds and energy errors are likely to increase the expected uncertainty in those parameters.  If the methods I used in Sec. \ref{sec:results} had failed for even ideal set of experiments, they would have certainly failed on the real deal.

There are two key features of current experiments that I keep.  First, I approximate the experimental efficiency $\mathcal{E}(Q)$ for each type of experiment to resemble those of current or recent experiments.  This efficiency is the probability that if there is a nuclear recoil of energy $Q$ somewhere in the experimental volume, it survives the selection cuts into the analysis.  The efficiency $\mathcal{E}(Q)$ includes both a fiducial volume cut as well as the acceptance probability within the fiducial volume.  I use the same efficiencies as used in Ref. \cite{peter2010b}.  Second, I retain the analysis windows (i.e., the nuclear recoil search window from the threshold energy \Qmin~to the maximum considered energy \Qmax) of current experiments, because as I show below in Sec. \ref{sec:results}, the analysis window strongly affects parameter estimation.  (\Qmin, \Qmax) is (2 keV, 30 keV) for the XENON1T-like experiment, (5 keV, 30 keV) for the LUX-like, (30 keV, 130 keV) for the argon experiment, and (10 keV, 100 keV) for the SuperCDMS-like experiment.

\subsection{Parameter estimation}\label{subsec:parameter}
Once I simulate mock data sets, I assess the parameter constraints using an unbinned likelihood function.  The probability that a single recoil is observed with energy $Q$ and with theoretical parameters $\{\theta\}$ and with experimental parameters (target nucleus, \Qmin, \Qmax, etc.) $\{\gamma\}$ is \cite{peter2010b}
\begin{eqnarray}
  P_{1}(Q|\{\theta\},\{\gamma\}) = \frac{\mathcal{E}(Q,\{\gamma\})dR/dQ(\{\theta\},\{\gamma\})}{\int_{Q_{\mathrm{min}}}^{Q_{\mathrm{max}}}dQ^\prime\mathcal{E}(Q^\prime,\{\gamma\})dR/dQ^\prime(\{\theta\},\{\gamma\})},
\end{eqnarray}
such that the likelihood of getting $N_e^i$ events of energy $\{Q^i_1,Q^i_2,...,Q^i_j\}$ in each experiment $i$ is
\begin{eqnarray}
  \mathcal{L}(\{Q\}|\{\theta\}) = \prod_{i=1}^{N} \frac{(N^i_e)^{N^i_o}e^{-N^i_e}}{N_o^i!} \prod_{j}^{N_o^i}P_{1}(Q_j^i|\{\theta\},\{\gamma_i\}),
\end{eqnarray}
where $N$ is the number of experiments and $N_o^{i}$ is the number of events observed in experiment $i$.  This form of the likelihood is currently used by both the CDMS and XENON100 experimental groups \cite{cdms2009,aprile2011}.

I use a Bayesian framework in which to determine the parameter uncertainties.  In this framework, the probability of the theoretical parameters of a given model hypothesis and the data, is
\begin{eqnarray}\label{eq:posterior}
  P(\{\theta\}|\{Q\}) \propto \mathcal{L}(\{Q\}| \{\theta\}) P(\{\theta\}),
\end{eqnarray}
which is also known as the posterior.  The coefficient relating the two sides of Eq. (\ref{eq:posterior}) is irrelevant for parameter estimation, so I replace ``$\propto$'' with ``$=$'' in that equation.   $P(\{\theta\})$ is the prior on the parameters.  I use the publicly-available {\sc MultiNest} nested sampling code to sample the posterior and determine parameter uncertainties \cite{feroz2008,feroz2009}.  For the results in Secs. \ref{subsec:1mb} and \ref{subsec:2mb}, I used 11000 live points for {\sc MultiNest}, and 16000 live points for the results in Sec. \ref{subsec:vel}.  For all the results discussed in Sec. \ref{sec:results}, I used a sampling efficiency of \texttt{efr}$=0.3$ and a tolerance on the accuracy of Bayesian evidence of \texttt{tol}$=10^{-4}$.  The values of \texttt{efr} and \texttt{tol} were chosen to get a good estimate of the maximum likelihood $\mathcal{L}_{\mathrm{max}}$ and the Bayesian evidence $\mathcal{Z}$.  The latter is the integral of the posterior over the volume of theoretical parameters
\begin{eqnarray}\label{eq:evidence}
  \mathcal{Z}(H|\{Q\}) = \int d\{\theta\} P(\{\theta\},\{Q\}).
\end{eqnarray}
Here, $H$ is the hypothesis for the model \cite{trotta2008}.  For example, a model hypothesis would be that all recoils in the direct-detection experiments are due to elastic scatters between WIMPs and nuclei and that the WIMP distribution function is described by a Maxwell-Boltzmann distribution.  Both $\mathcal{L}_{\mathrm{max}}$ and $\mathcal{Z}$ need to be calculated for Bayesian model-selection criteria, which I discuss in greater detail in Sec. \ref{subsec:vel}.

In addition, the low values of \texttt{efr} and \texttt{tol} allow one to estimate of the profile likelihood \cite{feroz2011}, which is defined as 
\begin{eqnarray}
\mathcal{L}_p(\{Q\}|\theta_i) = \max \left(\mathcal{L}(\{Q\}|\theta_i, \{\theta\})\right),
\end{eqnarray}
i.e., the maximum likelihood for a subset of the theoretical parameters fixed, over the space of the remaining theoretical parameters \cite{feroz2011}.  The profile likelihood is useful to calculate in addition to the marginalized posteriors to get a sense of whether the confidence limits based on the posterior are due to the size of the parameter space or due to high values of the likelihood.  See Refs. \cite{lewis2002,akrami2010,feroz2011} for more discussion.  In the following sections, I show confidence limits based on the marginalized posteriors and not the profile likelihood, using the latter as a sanity check.

The WIMP mass was sampled logarithmically in the interval $1\hbox{ MeV} < m_\chi < 100\hbox{ TeV}$, and the WIMP cross-section parameter $D = \rho_\chi \sigma_p^{SI}/m_\chi^2$ was sampled logarithmically from $10^{-60}\hbox{ GeV}^{-1}\hbox{cm}^{-1} < D < 10^{-40} \hbox{ GeV}^{-1}\hbox{cm}^{-1}$.  The speed-distribution parameters were sampled linearly, as described in Sec. \ref{sec:results}. 

It took {\sc MultiNest} approximately 4 CPU-hr to converge for each ensemble of mock data sets in Secs. \ref{subsec:1mb} and \ref{subsec:2mb} on a single processor on the University of California, Irvine's Greenplanet cluster, and from 18 to 150 CPU-hr for each ensemble in Sec. \ref{subsec:vel} depending on the dimensionality of the parameter space and the size of the data sets.  I found that the code slowed down dramatically if the number of parameters in the hypothesis exceeded $\sim 10$.

\section{Results}\label{sec:results}
In this section, I apply the analysis techniques in Sec. \ref{subsec:parameter} to mock data sets for several points in WIMP particle-physics and speed-distribution parameter space.

In Sec. \ref{subsec:1mb}, I estimate how well one may estimate \vlag~and \vrms~for single-Maxwell-Boltzmann distribution benchmark speed distributions of the form (\ref{eq:MB}) with a single-Maxwell-Boltzmann hypothesis.  Most forecasting studies have focused on a single benchmark speed distribution, but I show how the uncertainty on the WIMP mass, elastic scattering cross section, \vlag~and \vrms~depends sensitively on the underlying values of \vlag~and \vrms.  

In Sec. \ref{subsec:2mb}, I consider the case that the speed distribution is multimodal, but analyze the mock data sets with the hypothesis that the velocity distribution is Maxwell-Boltzmann.  The goal is to determine how biased the inferred WIMP mass and cross sections might be.  

In Sec. \ref{subsec:vel}, I analyze mock data with the SHM and Sec. \ref{subsec:2mb} multimodal benchmark speed distributions with the hypothesis that the speed distribution is a set of five step functions in geocentric speed.  This model of the speed distribution is supposed to be representative of a class of empirical models that may be used to fit the data.  While it is almost certainly not the optimal empirical hypothesis, it allows me to explore how well one may recover the WIMP mass, cross section, and speed distribution without a fixed, theoretically-inspired form for the DF.  In addition, I show that even for fairly small data sets, one may use Bayesian model-selection techniques to determine the relative quality of the fits for different hypotheses for the speed distribution.

In each section, I only consider one value of the parameter $D=\rho_\chi \sigma_p^{SI} /m_p^2$, setting $D=3\times 10^{-45}\hbox{ GeV}^{-1}\hbox{cm}^{-1}$.  I consider this parameter instead of treating $\sigma_p^{SI}$ and $\rho_\chi$ independently because of the total degeneracy of these parameters in direct-detection signals.  Only with outside information on \sigmapsi~(e.g., from future collider data sets) or $\rho_\chi$ may one place limits directly on the other parameter.  If one assumes $\rho_\chi = 0.3\hbox{ GeV cm}^{-3}$ \cite{kuijken1989b,bergstrom1998b}, then the fiducial value of $D$ implies $\sigma_p^{SI} \approx 10^{-44}\hbox{ cm}^2$, which is a factor of several below the minimum of the current $m_\chi-\sigma_p^{SI}$ exclusion curve.  This value of $D$ should be accessible to next-generation direct-detection experiments.  Note, though, that the exclusion curve is constructed by fixing the WIMP speed distribution to a particular model.  

In both Secs. \ref{subsec:1mb} and \ref{subsec:2mb}, there are four free parameters to fit: \mwimp, $D$, \vlag, and \vrms.  In Sec. \ref{subsec:vel}, the number of free parameters is two plus the number of step functions used to describe the speed distribution.

\subsection{Single Maxwell-Boltzmann Distribution, In and Out}\label{subsec:1mb}
The first test is to see how well one may infer WIMP particle-physics and speed-distribution parameters in the case that the hypothesis for the form of the speed distribution matches the form of the true distribution.  In particular, I focus on parameter constraints for the benchmark SHM and variations to it, making the most minimal of prior assumptions about any of the parameters of the WIMP and Maxwell-Boltzmann model hypothesis: $\{\theta\} = \{m_\chi, D, v_{\mathrm{lag}},v_{\mathrm{rms}}\}$.  While previous forecasting studies have considered a variety of benchmark \mwimp, nearly all (with the exception of Refs. \cite{peter2010b,green2010}) have considered only one fiducial speed distribution with fixed \vlag~and \vrms.  However, even with the ansatz that the local WIMP density is dominated by a smooth, equilibrium halo component (neglecting the accretion-history-dependent features seen in high-resolution N-body simulations and any anisotropy in the velocity ellipsoid \cite{vogelsberger2009,kuhlen2010}) with one of the theoretically-inspired forms of the speed distribution, there is still a great deal of uncertainty on the appropriate values of \vlag~and \vrms~for the Milky Way.

With the ansatz that the local WIMP DF results from a smooth, equilibrium, nonrotating dark-matter halo DF, the appropriate choice for \vlag~is the sum of the velocity of the LSR \cite{binney2008}, solar motion (the peculiar speed of the Sun relative to the LSR) \cite{schoenrich2010}, and the velocity of the Earth about the Sun.  The largest uncertainty on any of those components is on the LSR.  While the IAU standard is $v_{\mathrm{LSR}} = 220 \hbox{ km s}^{-1}$ with approximately 10\% uncertainty \cite{kerr1986}, more recent measurements of the rotation curve and of the mass of the Milky Way halo indicate that slightly larger values are preferred \cite{reid2009,watkins2010}.  However, the uncertainty in the speed of the LSR from any measurement in the past several decades has not changed (see, e.g., Ref. \cite{scott2008}), so the range of plausibility for the speed of the LSR is ~$200-270\hbox{ km s}^{-1}$.  With the addition of solar motion and the velocity of the Earth about the Sun, in this work I consider the range of plausibility for \vlag~to be $220-280\hbox{ km s}^{-1}$.

It is not clear what the best choice for \vrms~is.  For a power-law dark-matter density profile $\rho(r) \propto r^{-\beta}$ and a flat rotation curve, it can be shown that the distribution function is Maxwell-Boltzmann velocity dispersion 
\begin{eqnarray}\label{eq:beta}
v_{\mathrm{rms}} = v_{\mathrm{lag}}/\sqrt{\beta}
\end{eqnarray}
if one assumes that the velocity ellipsoid is isotropic \cite{peter2008}.  If dark-matter profiles are described by a Navarro-Frenk-White density profile with a scale radius $r_s$, then $\rho(r) \propto r^{-1}$ for $r \ll r_s$, $\rho(r) \propto r^{-2}$ for $r\sim r_s$, and $\rho(r) \propto r^{-3}$ \cite{navarro1996,navarro1997}.  Neglecting the effects of baryons on dark-matter halos, for a dark-matter halo of mass $(1-3)\times 10^{12}M_\odot$ (the plausible range of values for the Milky Way's virial mass \cite{xue2008,reid2009,watkins2010}), the typical scale radius should be of order $10-30$ kpc \cite{bullock2001}.  Given that the Sun sits $\sim 8$ kpc from the Galactic center \cite{ghez2008}, it is plausible that $\beta \sim 1-2$.

\begin{figure*}[p]
\centering
\includegraphics[width=0.61\textwidth]{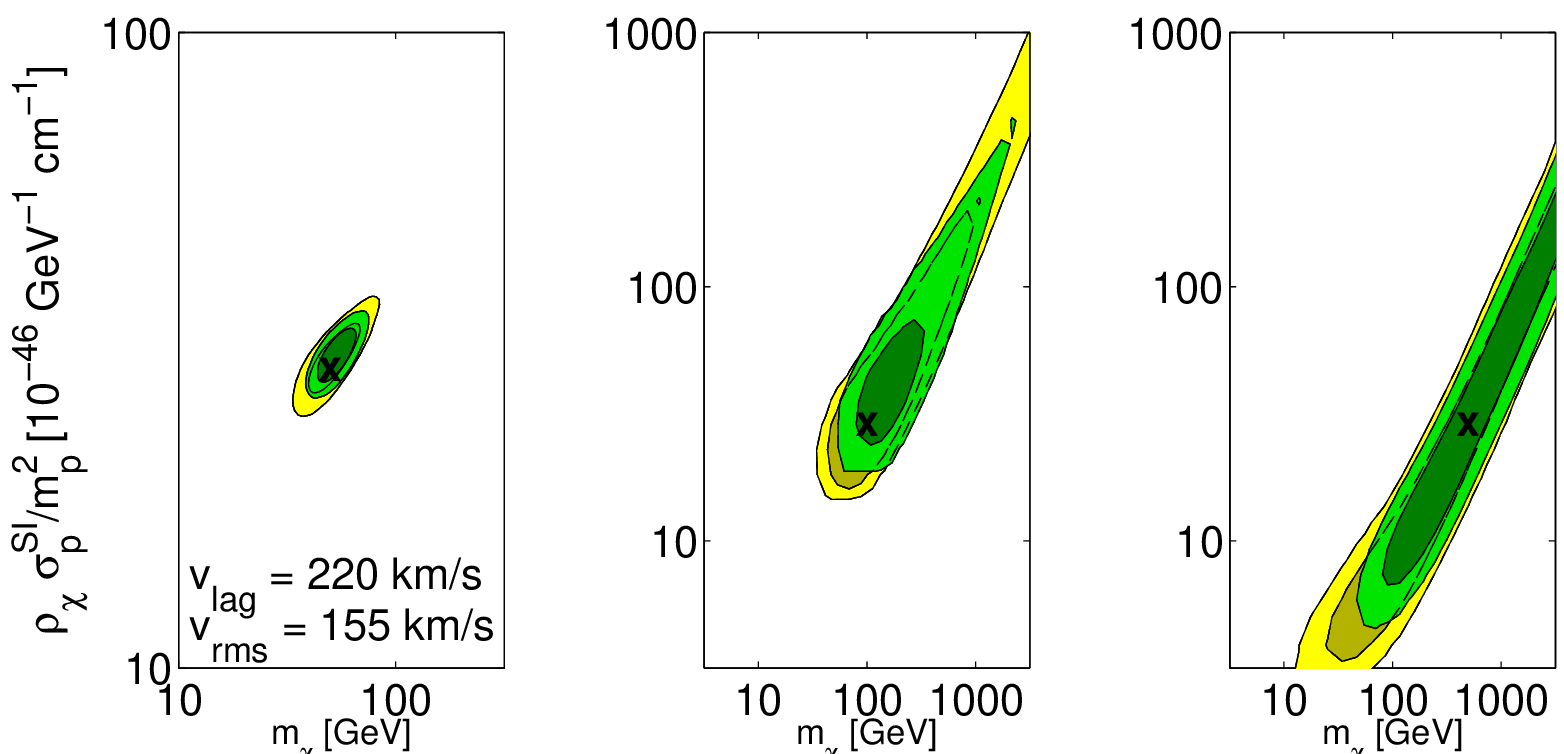}\\
\includegraphics[width=0.61\textwidth]{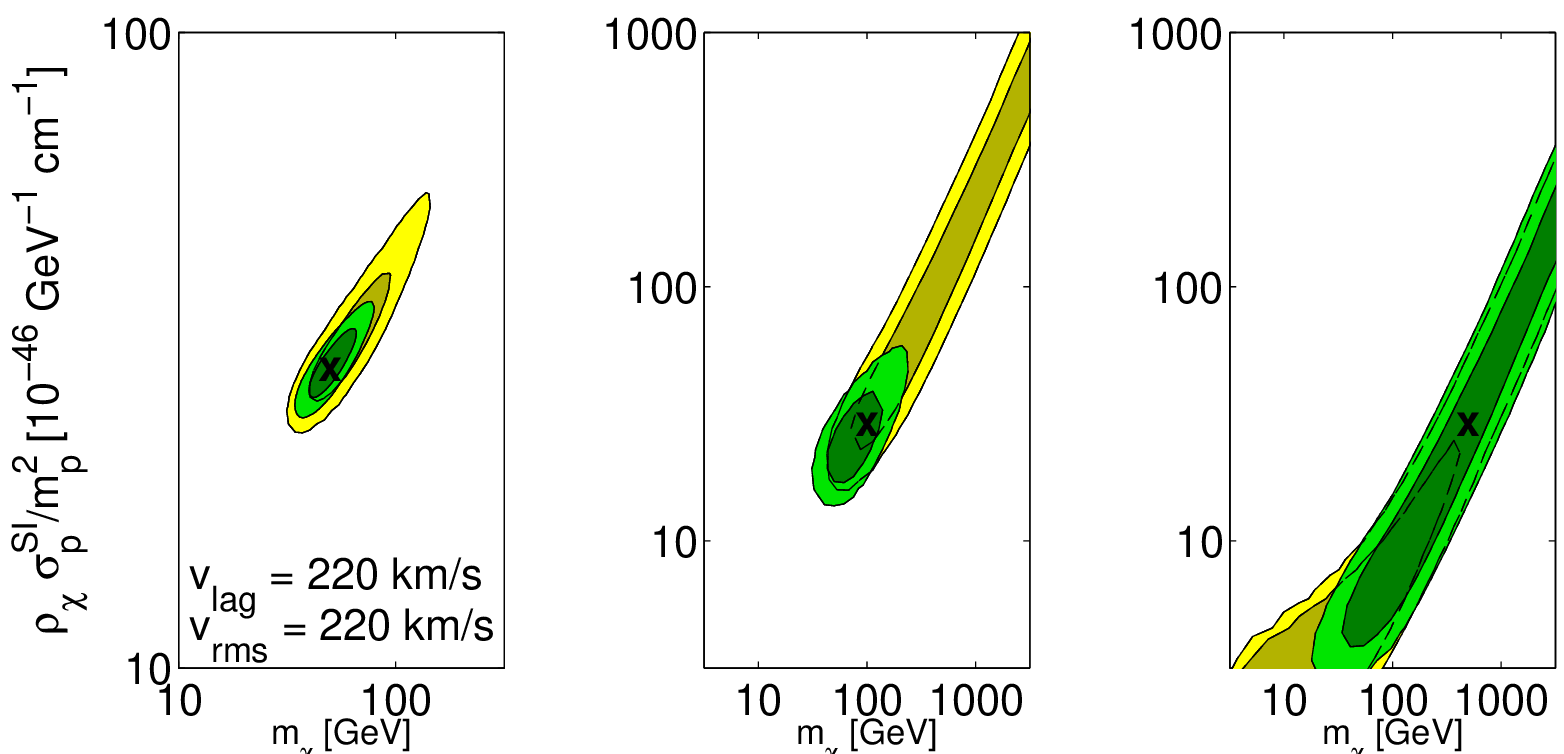}\\
\includegraphics[width=0.61\textwidth]{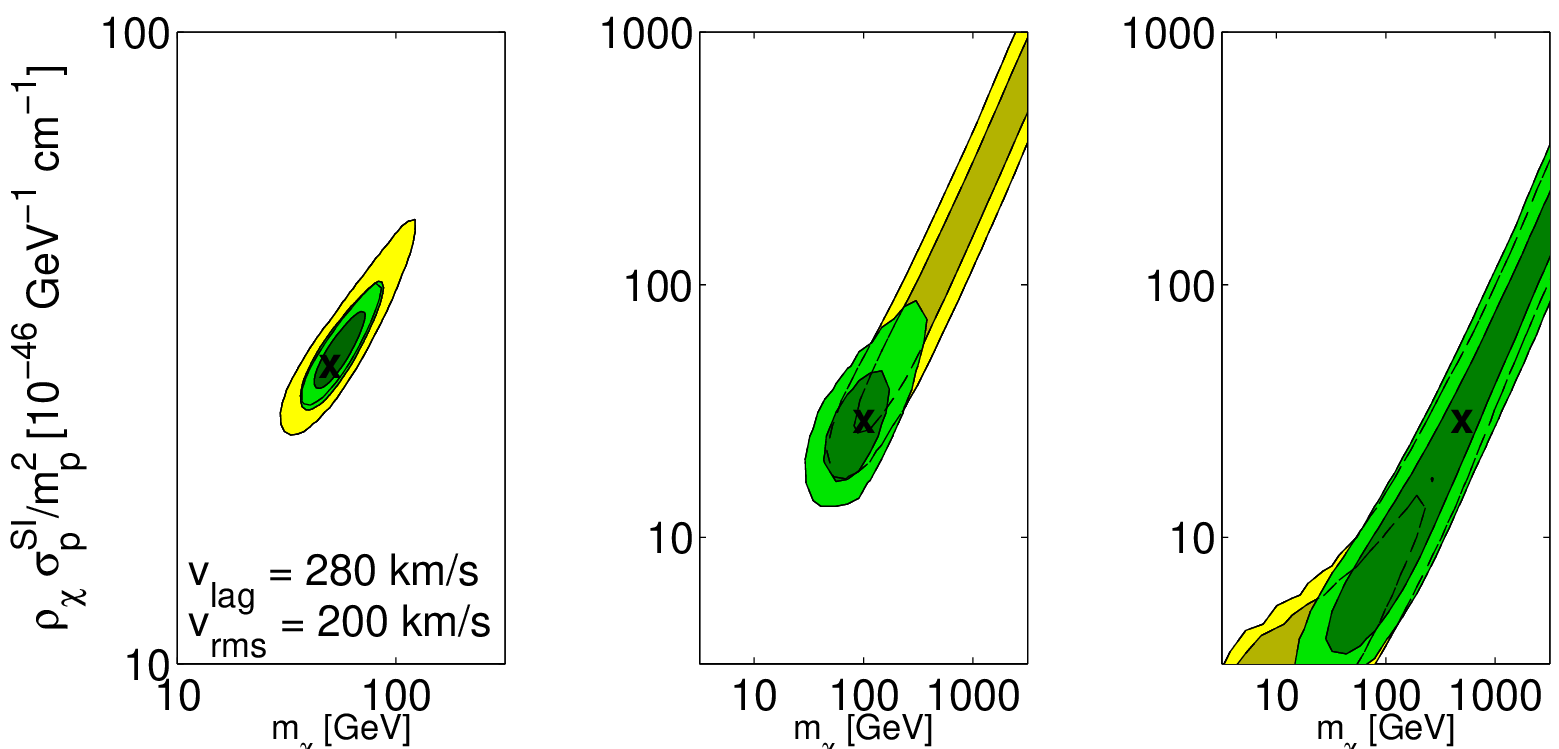}\\
\includegraphics[width=0.61\textwidth]{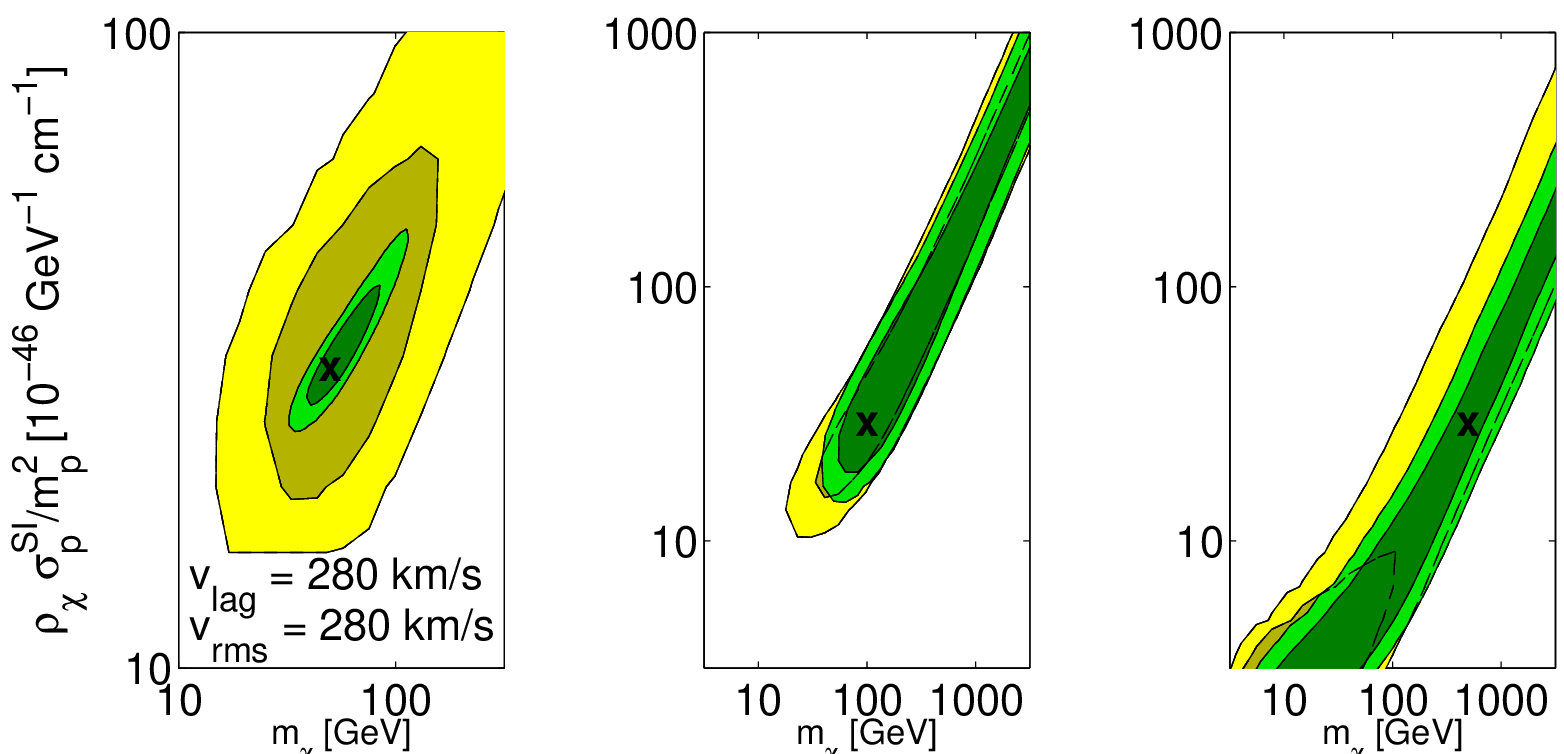}
\caption{\label{fig:oneg_msigma}Marginalized probability distributions for \mwimp~and $D = \rho_\chi \sigma_P^{SI}/m_p^2$.  The 68\% C.L. region is darker than the 95\% C.L. region.  The lighter pair of contours is associated with WIMP searches in the analysis windows described in Sec. \ref{subsec:toy}, and the darker pair is associated with extending the analysis window to 1 MeV.  Each row of figures corresponds to a different WIMP speed-distribution benchmark model.  For each, a single Maxwell-Boltzmann velocity distribution is assumed (of the form in Eq. (\ref{eq:MB})), but with different \vlag~and \vrms.  The x's mark the input \mwimp~and \sigmapsi assuming $\rho_\chi = 0.3\hbox{ GeV cm}^{-3}$.}
\end{figure*}

The first step in this analysis is to see how the constraints on \mwimp~and $D$ are affected by the underlying WIMP speed distribution.  I consider three benchmark WIMP masses: \mwimp$=50,100,$ and 500 GeV.  I bracket the range of plausible \vlag~and \vrms~with the following benchmark Maxwell-Boltzmann DFs: the SHM; \vlag$=220\hbox{ km s}^{-1}$ and \vrms$=220\hbox{ km s}^{-1}$; \vlag$=280\hbox{ km s}^{-1}$ and \vrms$=200\hbox{ km s}^{-1}$; and \vlag$=280\hbox{ km s}^{-1}$ and \vrms$=280\hbox{ km s}^{-1}$.  The mock data sets had of order 100 events for the LUX-like experiment, of order tens to a hundred events for the  argon experiment, of order ten or tens for the SuperCDMS-like experiment, and several tens to hundreds of events for the XENON1T-like experiment.  The latter has a relatively high number of events due to the low energy threshold \Qmin$=2$ keV.  The total number of events in all toy experiments decreased with increasing WIMP mass due to the fact that the number density of WIMPs $n_\chi \propto m_\chi^{-1}$ and that the typical WIMP speeds were high enough that there were many events above threshold for all experiments.  

As described in Sec. \ref{subsec:parameter}, I sampled $D$ and \mwimp~logarithmically for the {\sc MultiNest} nested sampler.  I sampled \vlag~and \vrms~linearly in the range $0-2000\hbox{ km s}^{-1}$.  Even though this range is far broader than the ``plausible'' ranges for these parameters, I want to explore parameter constraints with weak priors.  If the data are sufficiently good, the parameter constraints should depend little on the prior.  Since my choice of $D$ is somewhat optimistic, if the parameter constraints are prior dependent for even this value of $D$, then parameter inference for 2015-era direct-detection experiments will be heavily prior dependent.  The upper end of the range for \vlag~and \vrms~is far above the current best estimates for the local escape speed from the Galaxy, $v_{\mathrm{esc}} \approx 550 \hbox{ km s}^{-1}$ \cite{smith2007}.

The reconstructed \mwimp~and $D$ are shown with the light-color-filled contours in Fig. \ref{fig:oneg_msigma}, panels of which were made using a modified version of the publicly-available {\sc cosmomc} \texttt{getdist} code \cite{lewis2002}.  Each column in the figure shows the results for a single WIMP mass, and each row shows a different speed-distribution benchmark.  The 68\% and 95\% confidence limits (C.L.) are actually central credible intervals, for which equal volumes of the posterior lie outside the upper and lower edges of the intervals \cite{hamann2007}.  This is how the C.L.'s will be defined for the rest of this work.  Generically, it is possible to get good constraints on low-mass WIMPs even without strong priors on the speed-distribution parameters \vlag~and \vrms, although the constraints for $m_\chi = 50$ GeV are much tighter for the SHM that the other equilibrium halo models.  However, the constraints if $m_\chi = 50$ GeV are poor if \vlag$=$\vrms$=280\hbox{ km s}^{-1}$.

\begin{figure*}[t]
\centering
\includegraphics[width=0.6\textwidth]{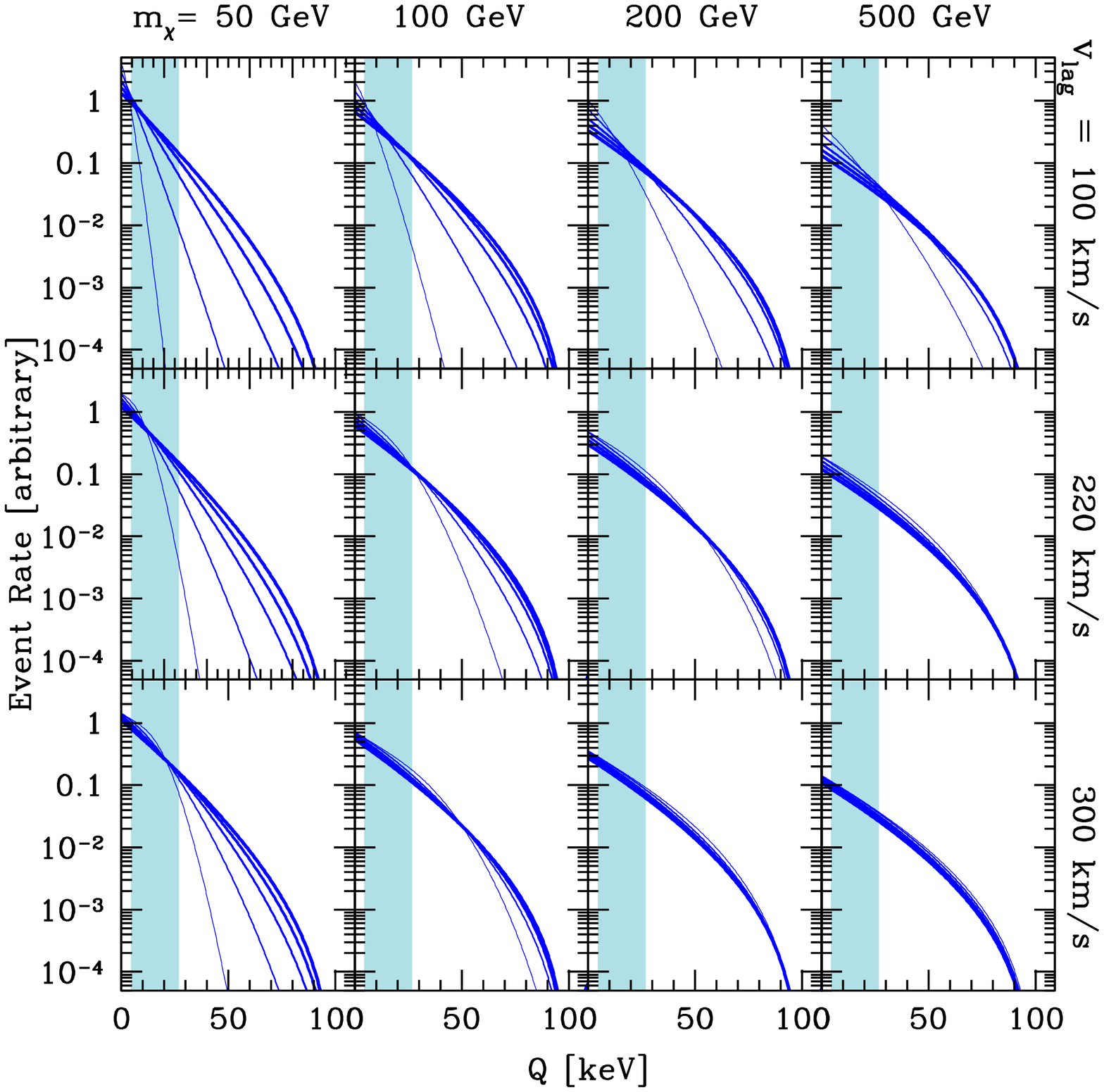}
\caption{\label{fig:drdq_xe}Recoil energy spectra for a xenon-based experiment for different \mwimp, \vlag, and \vrms.  Each column represents a different \mwimp, and each row represents a different \vlag.  The lines on the plots represent different \vrms, with (thinnest to thickest) \vrms$=50,100,150,200,250\hbox{ km s}^{-1}$.  The shaded region shows the XENON10 analysis window \cite{angle2008}.
}
\end{figure*}

 The constraints for $m_\chi \gtrsim 100$ GeV are much poorer than for $m_\chi \lesssim 50$ GeV.  In general, it is only possible to find a lower limit for the WIMP mass and cross section.  This is because the typical recoil energy is $Q \sim \mu_A v_{\mathrm{lag}}^2/m_A$, where $\mu_A$ is the reduced mass for the WIMP-nucleon system.  For $m_\chi / m_A \gg 1$, $\mu_A \rightarrow m_A$.  Thus, the recoil spectrum is independent of WIMP mass for sufficiently high-mass WIMPs.  This point is illustrated in Fig. \ref{fig:drdq_xe}, in which I plot recoil spectra of xenon as a function of \mwimp, \vlag, and \vrms.  Each column represents a different WIMP mass, each row a different \vlag, and the line thickness signifies the value of \vrms.  The shaded region is the analysis window for the XENON10 experiment, the analysis window I use as the default for the LUX-like toy experiment \cite{angle2008}.

The shapes of the recoil spectra in and outside of the analysis window in Fig. \ref{fig:drdq_xe} indicate a possible way to more tightly constrain the WIMP parameters: extend the analysis windows to higher energy.  A larger analysis window gives one a longer lever arm on the recoil spectrum.  In Fig. \ref{fig:drdq_xe}, there are a number of recoil spectra that look nearly identical inside the analysis window but diverge outside.  Even a few recorded events at high recoil energy could prove useful in parameter constraints.  The darker set of contours in Fig. \ref{fig:oneg_msigma} indicate parameter constraints when the upper end of the analysis window is extended to \Qmax$= 1$ MeV for all experiments.  There is only a modest increase in the number of events relative to the number of events in the fiducial analysis windows ($\sim 5\%-25\%$ depending on the WIMP mass, target nucleus, and speed distribution), but the constraints in the \mwimp$-D$ plane are obviously significantly better, especially for the \mwimp$=50$ GeV and 100 GeV cases.  The question is if backgrounds at higher energies will limit the constraining power of the high-energy nuclear recoils.  I defer that subject to future work.

\begin{figure*}[p]
\centering
\includegraphics[width=0.61\textwidth]{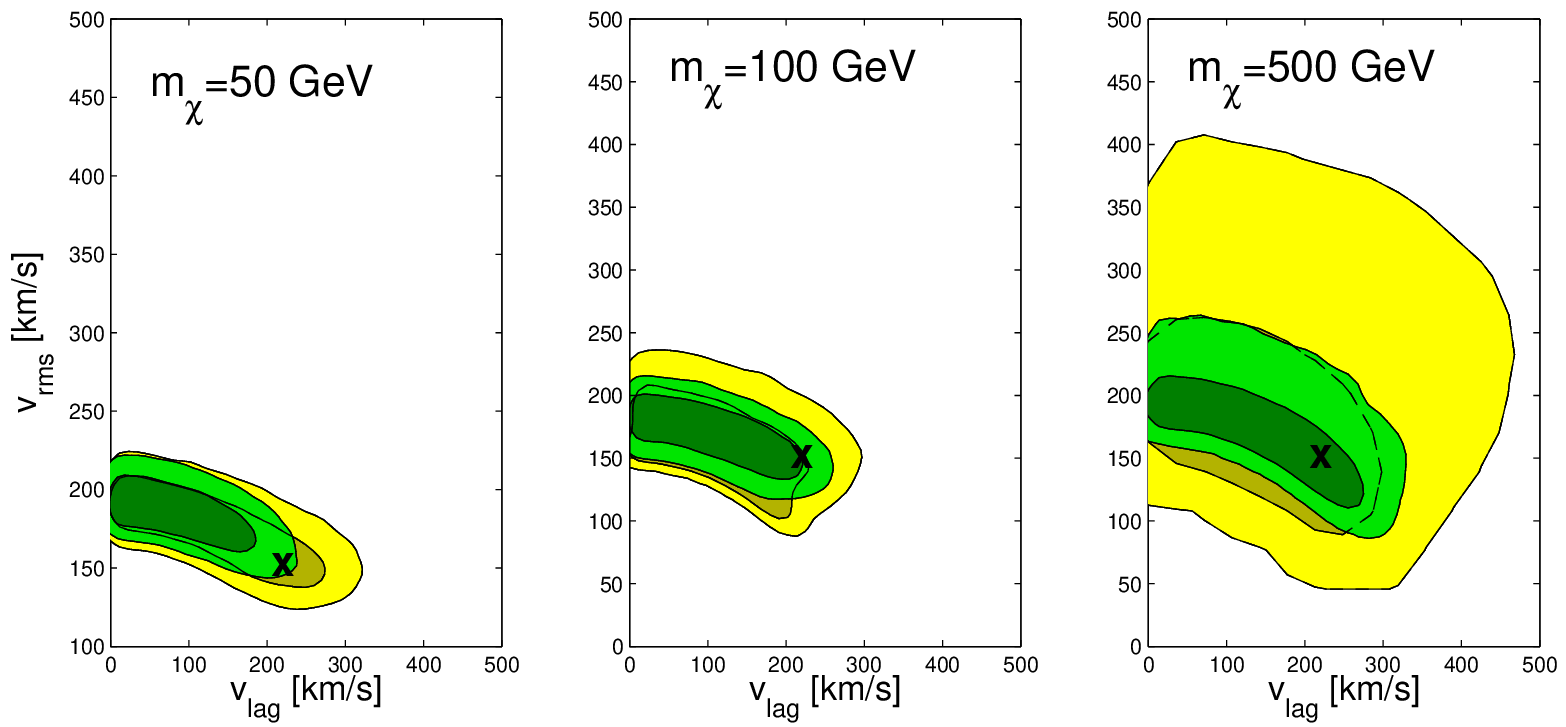}\\
\includegraphics[width=0.61\textwidth]{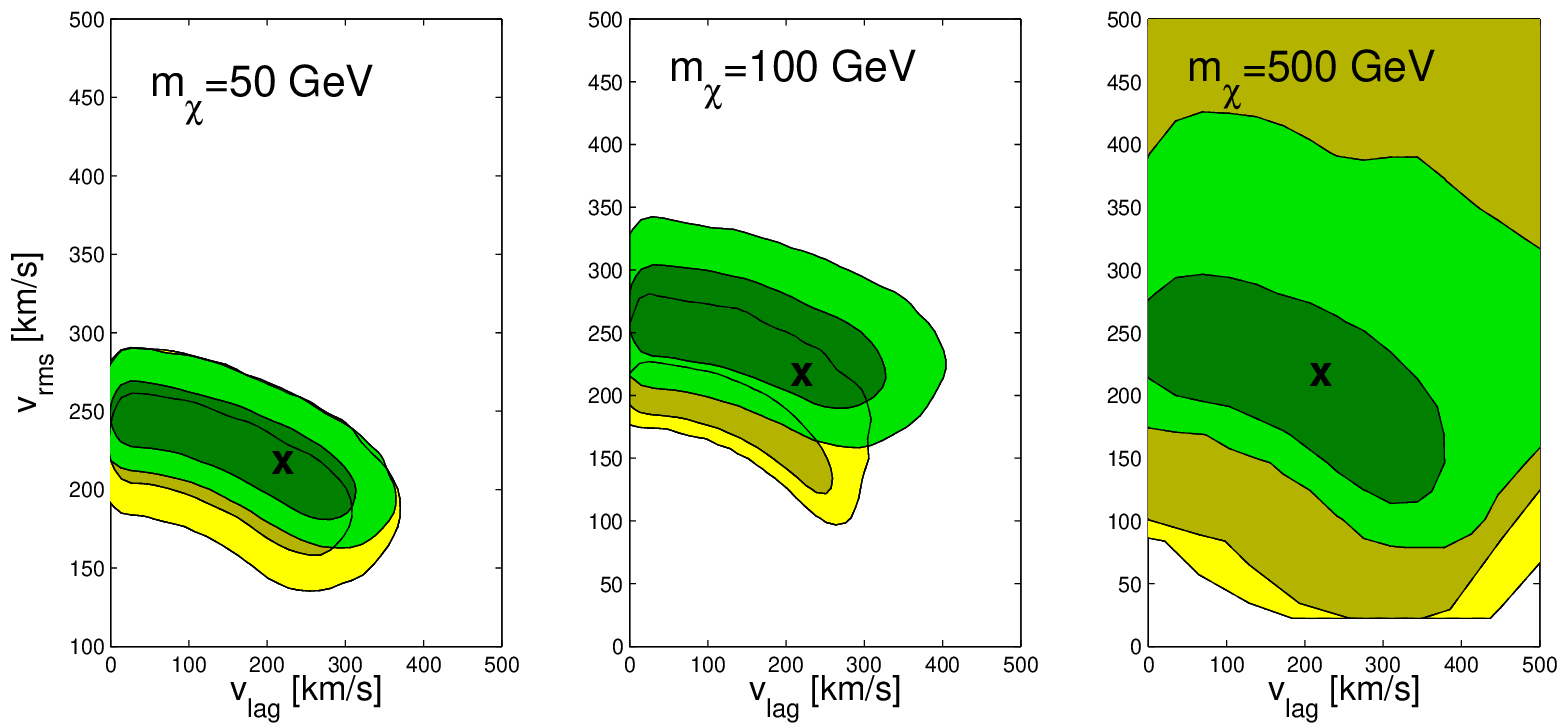}\\
\includegraphics[width=0.61\textwidth]{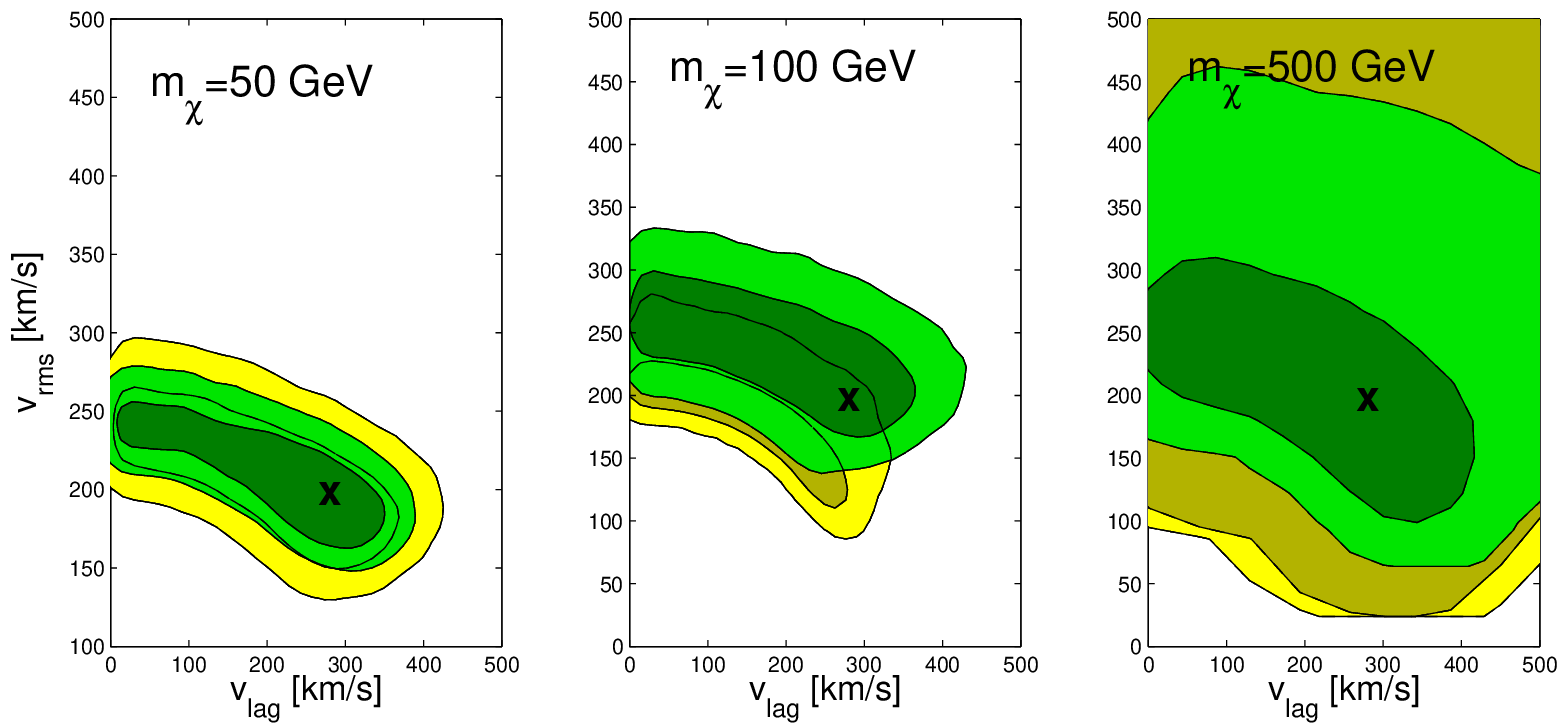}\\
\includegraphics[width=0.61\textwidth]{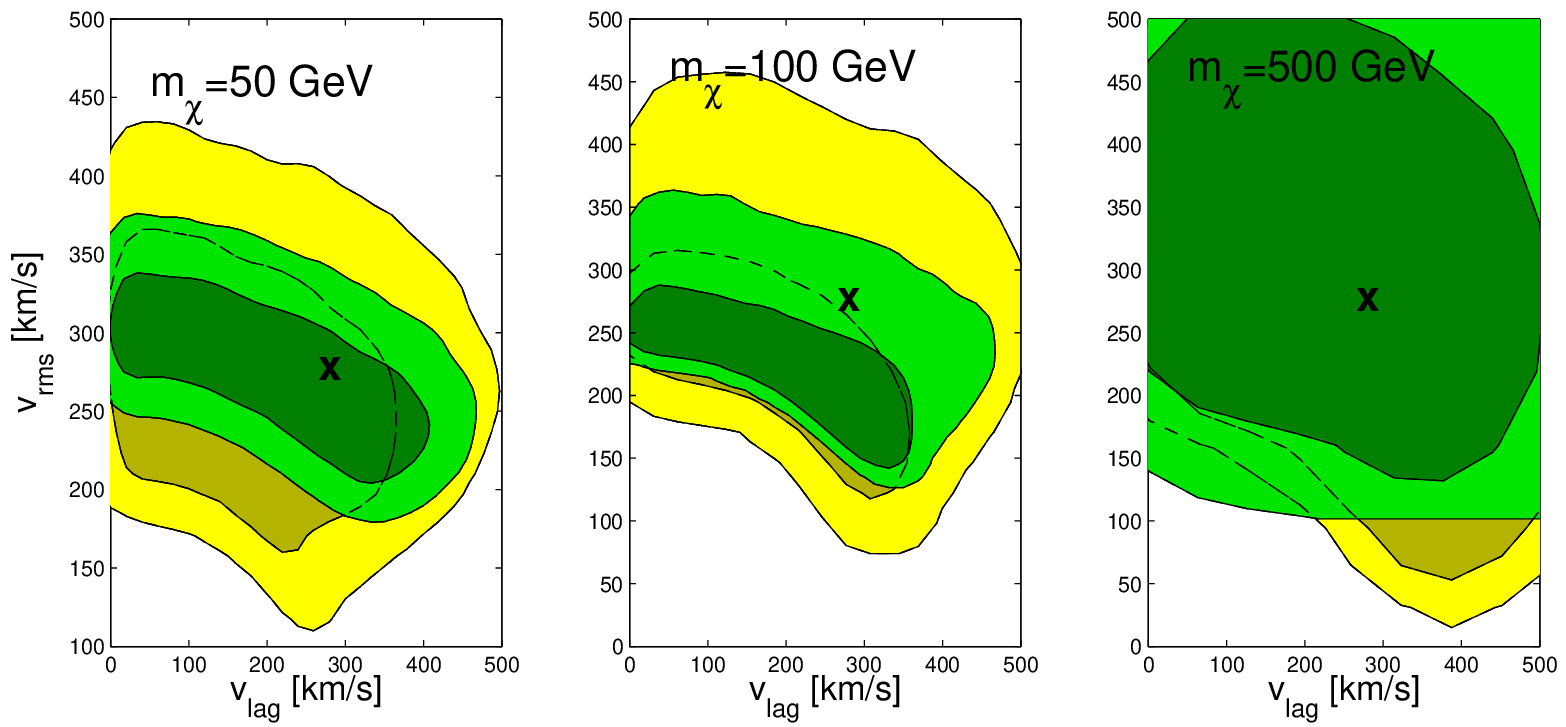}
\caption{\label{fig:oneg_v}Marginalized probability distributions for \vlag~and \vrms~assuming a single Maxwellian speed distribution.  Contours show 68\% and 95\% C.L regions.  The lighter pair of contours is associated with WIMP searches in the analysis windows described in Sec. \ref{subsec:toy}, and the darker pair are associated with extending the analysis window to 1 MeV.  Each row of figures corresponds to a different benchmark WIMP velocity model with (top to bottom): \vlag$=220\hbox{ km s}^{-1}$ \& \vrms$=155\hbox{ km s}^{-1}$; \vlag$=220\hbox{ km s}^{-1}$ \& \vrms$=220\hbox{ km s}^{-1}$; \vlag$=280\hbox{ km s}^{-1}$ \& \vrms$=200\hbox{ km s}^{-1}$; and \vlag$=280\hbox{ km s}^{-1}$ \& \vrms$=280\hbox{ km s}^{-1}$.  Each column represents a different benchmark WIMP mass.}
\end{figure*}

Next, I examine the constraints on the WIMP speed-distribution parameters \vlag~and \vrms, which are shown in Fig. \ref{fig:oneg_v}.  As in Fig. \ref{fig:oneg_msigma}, the lighter set of marginalized probability contours corresponds to the fiducial analysis windows, and the darker set of contours corresponds to the analysis in which \Qmax~is increased to 1 MeV.  As in the \mwimp$-D$ plane, the constraints are tighter for higher \Qmax.  The speed-distribution constraints are generally better for low-mass WIMPs than for \mwimp$=500$ GeV.  For the lower-mass WIMPs, there is a long tail in the posterior towards small \vlag.  This has to do with the fact that although the typical WIMP speed \vlag~is important in setting the typical energy scale of the events, the distribution of speeds \vrms~governs the shape of the recoil spectrum.  For example, if the WIMP distribution function were a delta function centered on \vlag (the limit of infinitely small \vrms), the recoil spectrum divided through by $F^2_{SI}(Q)$ would be a step function that cuts off when $v_\mathrm{min}$ exceeds \vlag.  If, however, the distribution function were flat up to some cut-off such that the typical speed were \vlag~(the limit of large \vrms), there would be a longer tail in the recoil spectrum to higher $Q$ since this distribution would have a number of high-speed WIMPs.

There are already two conclusions we can draw from this study.  First, one may simultaneously constrain the parameters of the model (\mwimp, $D$, \vlag, and~\vrms) from direct-detection data without strong priors on any of those parameters, assuming that the true WIMP DF looks something like a Maxwell-Boltzmann distribution and there are at least of order 100 events in all experiments combined.  The constraints do vary as a function of all those parameters, and appear best for the SHM versus halo models with higher \vlag~and \vrms.  Second, the constraints improve significantly if the analysis window is extended to higher energies, at least if backgrounds are negligible.

\begin{figure*}[p]
\centering
\includegraphics[width=0.605\textwidth]{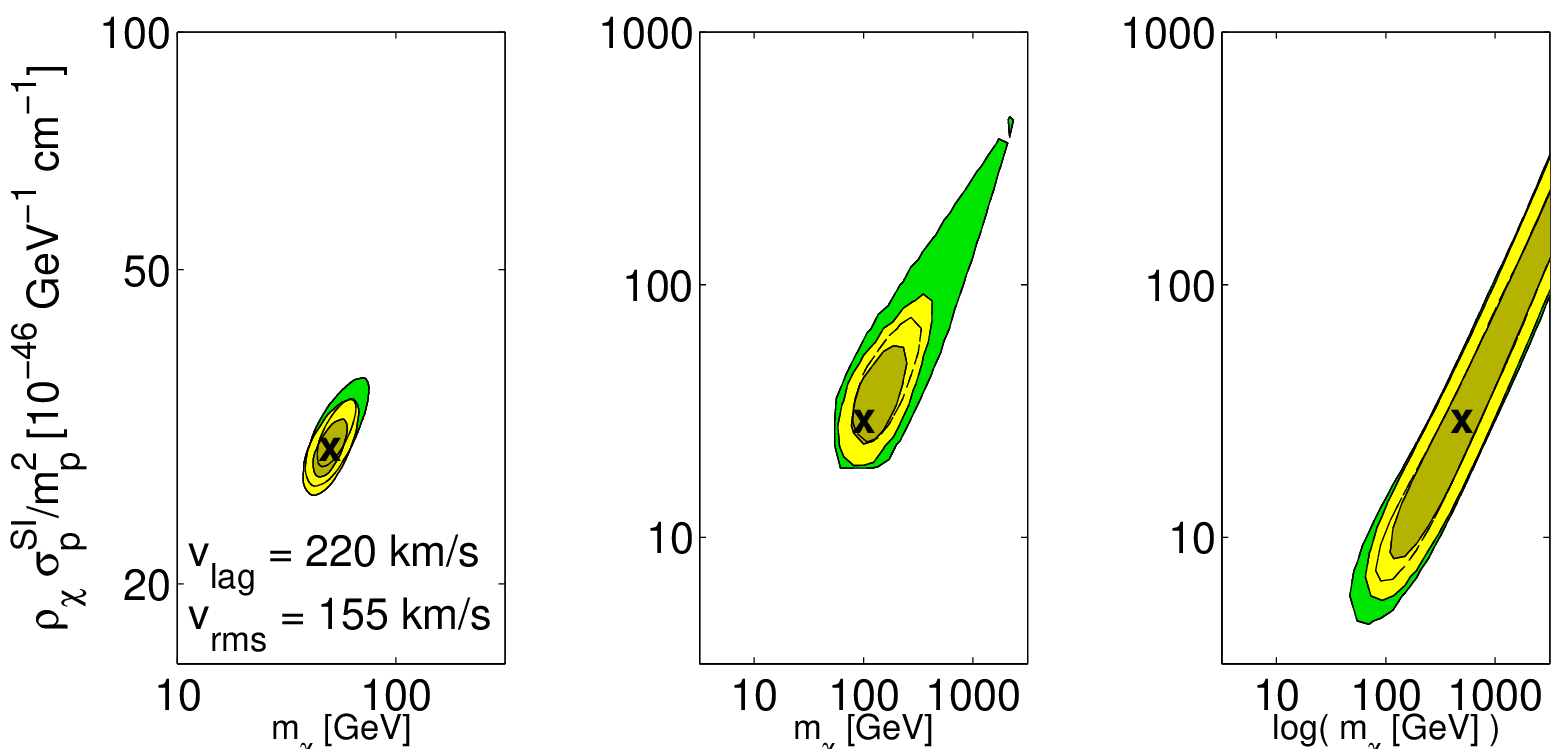}\\
\includegraphics[width=0.605\textwidth]{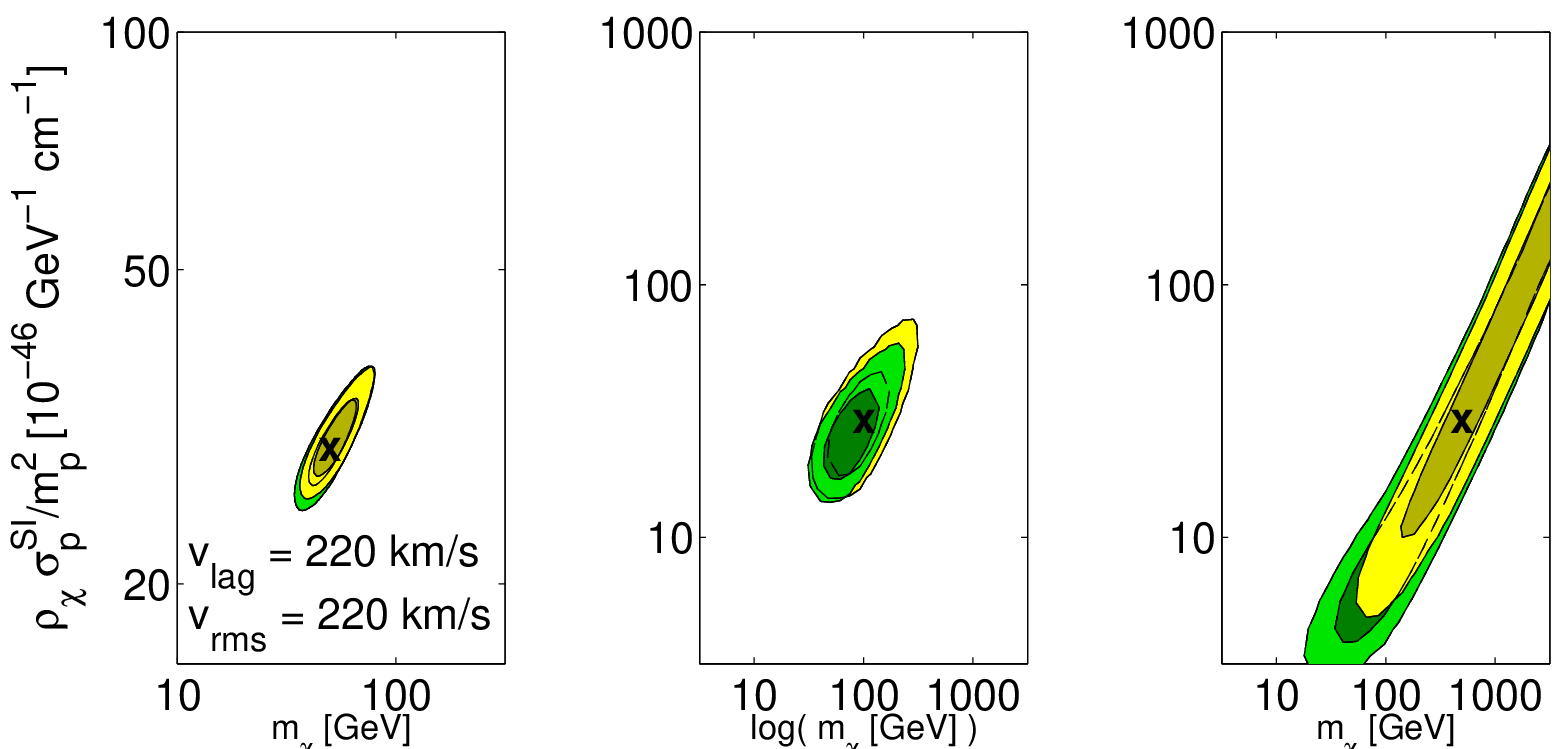}\\
\includegraphics[width=0.605\textwidth]{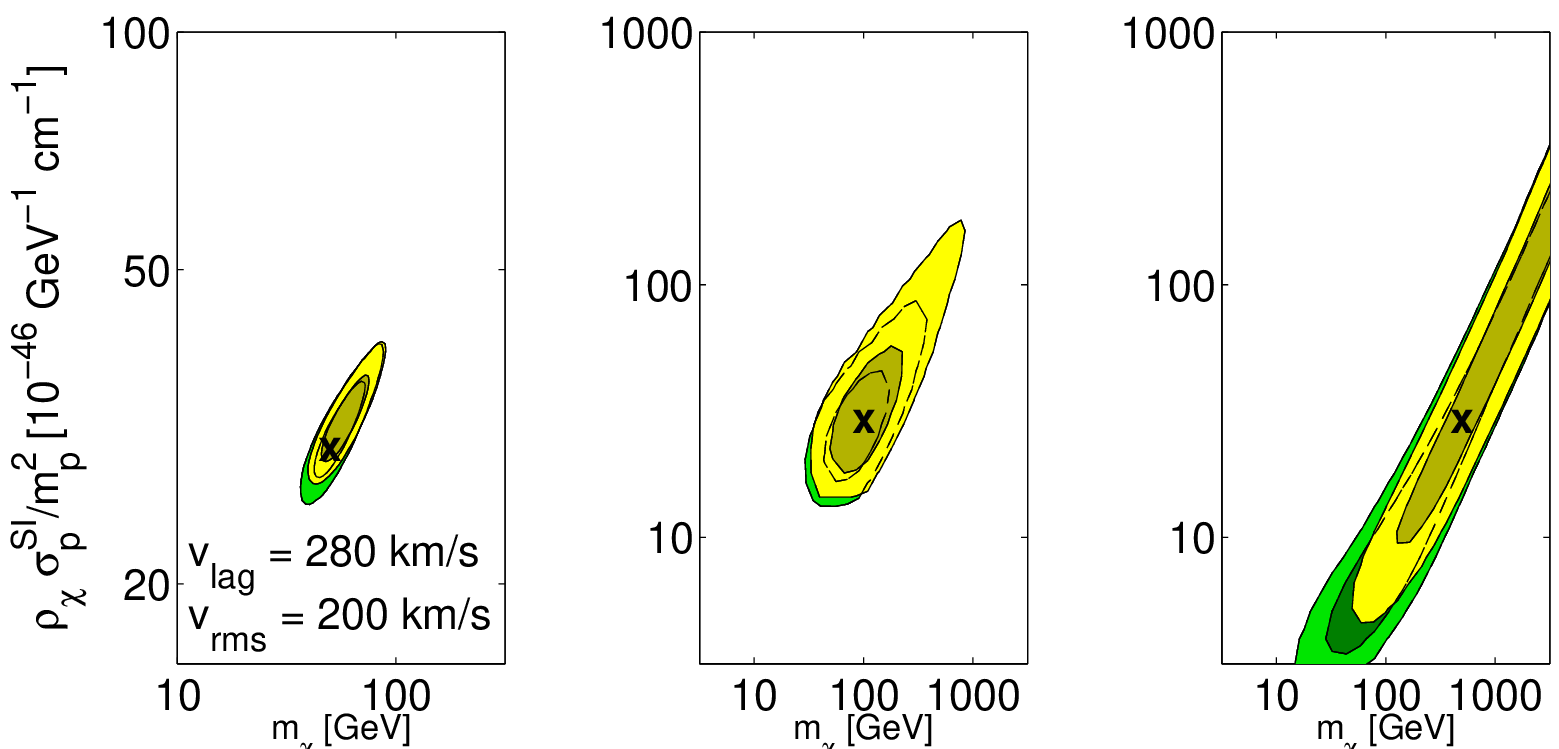}\\
\includegraphics[width=0.605\textwidth]{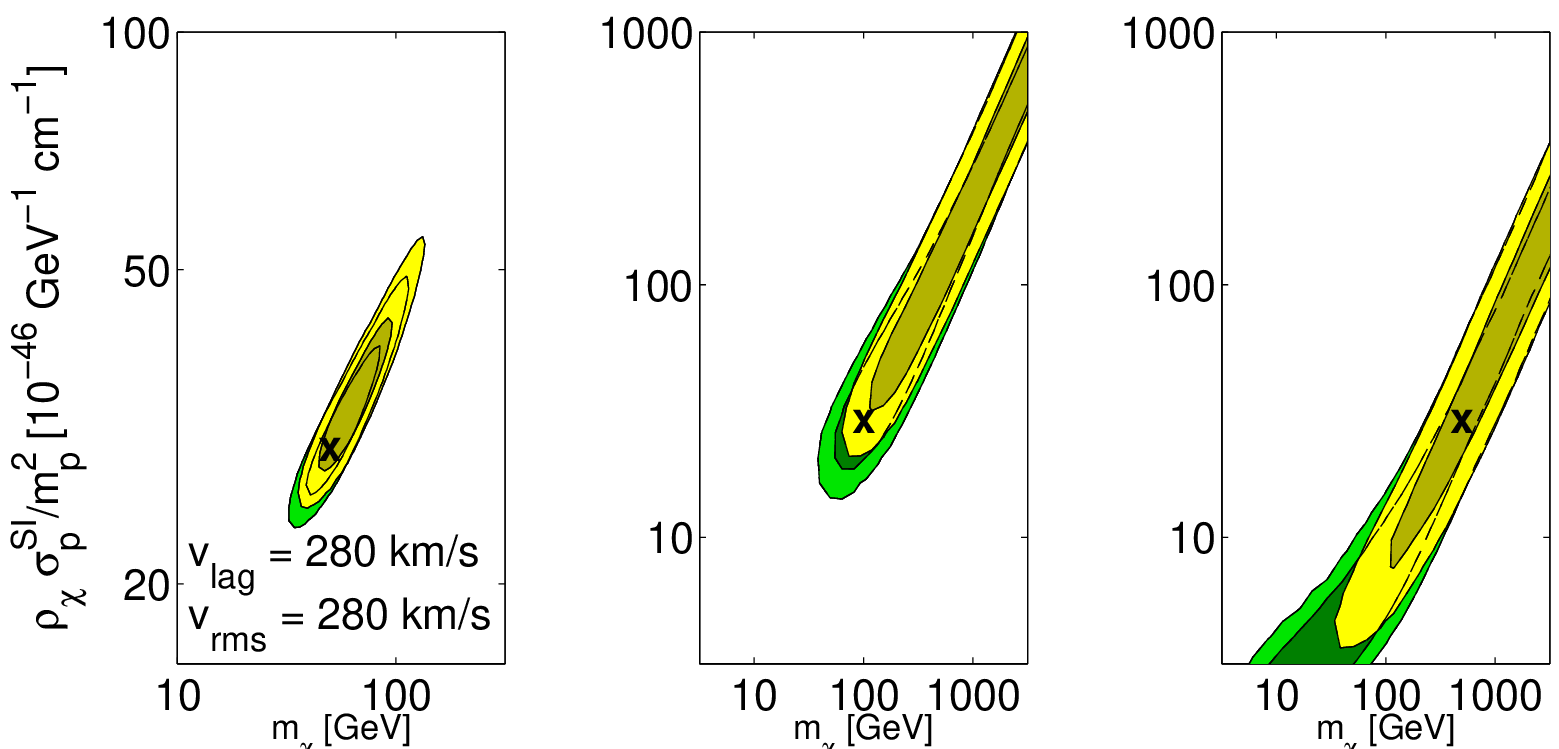}
\caption{\label{fig:oneg_msigma_prior7}Marginalized probability distributions for \mwimp~and $D = \rho_\chi \sigma_P^{SI}/m_p^2$~with Maxwellian speed distributions for the WIMPs as the benchmark models and as the hypothesis.  The lines outline 68\% and 95\% C.L. regions.  The darker pair of contours is associated with WIMP searches with flat priors on the velocity parameters with $Q_{\mathrm{max}} = 1$ MeV for all experiments, and the lighter pair are associated with a 10\% Gaussian prior on $v_{\mathrm{lag}} = 220\hbox{ km s}^{-1}$ and a prior on $\beta$ (Eq. \ref{eq:beta}; see text for details).  Each row of figures corresponds to a different benchmark WIMP velocity distribution, and each column represents a different benchmark WIMP mass.}
\end{figure*}

It is useful to see how the constraints on \mwimp~and $D$ compare to the case in which strong priors are placed on the speed distribution, as is typically done in WIMP parameter forecasts \cite{green2008a,pato2010}.  I consider a prior that consists of a Gaussian for \vlag~centered at 220$\hbox{ km s}^{-1}$ and with a width of 22$\hbox{ km s}^{-1}$ multiplied by a Gaussian prior on $\beta$ centered on $\sqrt{2}$ with a width of 0.4.  The width of the prior on \vlag~is the IAU value of the uncertainty on the speed of the LSR, and the prior on $\beta$ spans the values expected for a Navarro-Frenk-White profile.  In Fig. \ref{fig:oneg_msigma_prior7}, I show the constraints in the \mwimp$-D$ plane using \Qmax$=1$ MeV with this prior (light-colored filled contours) and the constraints without the prior with the same \Qmax.  The constraints on \mwimp~and $D$ are not significantly better with the inclusion of the strong prior---the data are sufficient for the likelihood to influence the posterior away from the prior, although not entirely.  This also means that the strong prior does not significantly bias the constraints on \mwimp~and $D$.  The only case in which the prior does somewhat improve the fit is for the SHM, which is unsurprising because the priors are centered on SHM parameters.  The takeaway message from Fig. \ref{fig:oneg_msigma_prior7} is that imposing a strong prior on the speed-distribution parameters is unnecessary, at least if there are at least of order 100 events in all experiments combined.  If there are fewer events, the parameter constraints may be prior dominated if a strong prior is imposed.

\begin{figure*}[p]
\centering
\includegraphics[width=0.61\textwidth]{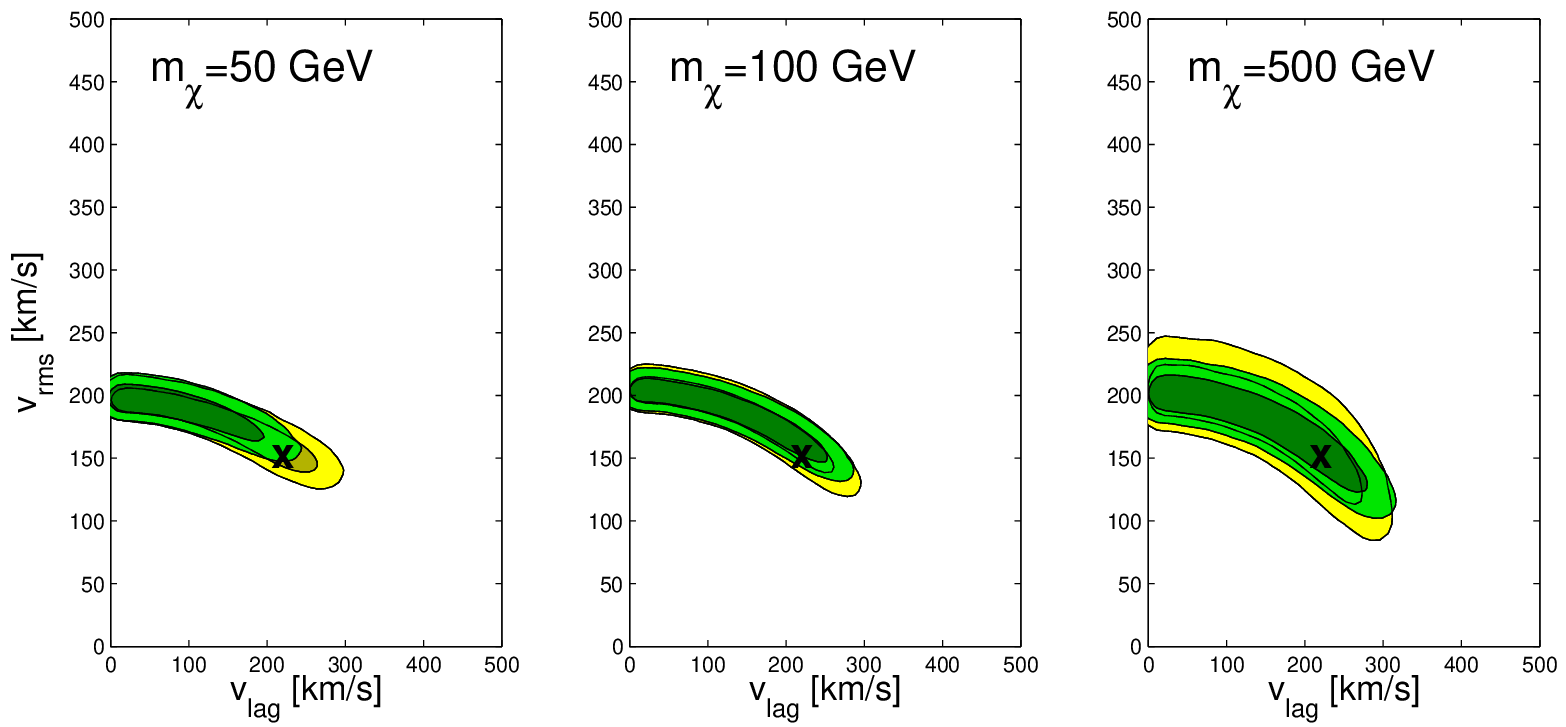}\\
\includegraphics[width=0.61\textwidth]{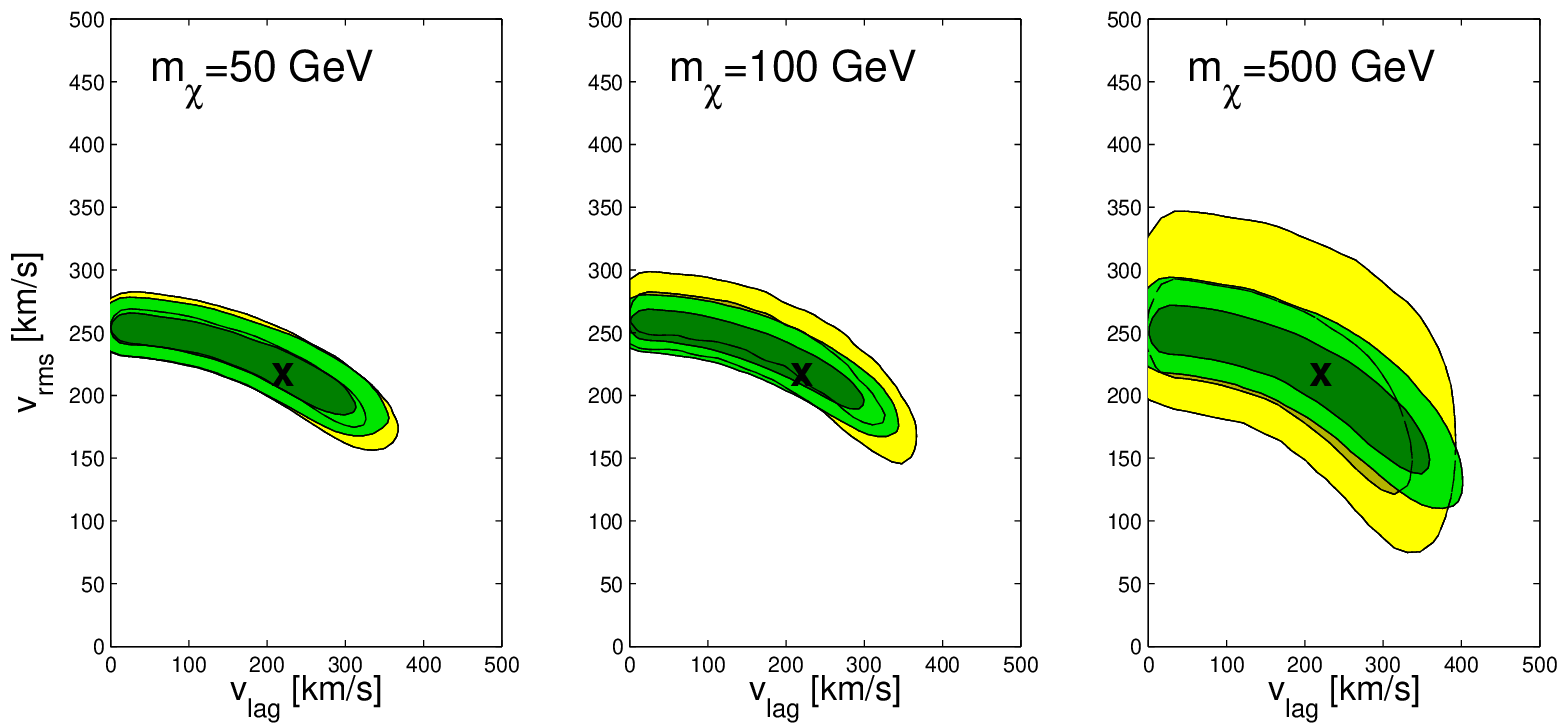}\\
\includegraphics[width=0.61\textwidth]{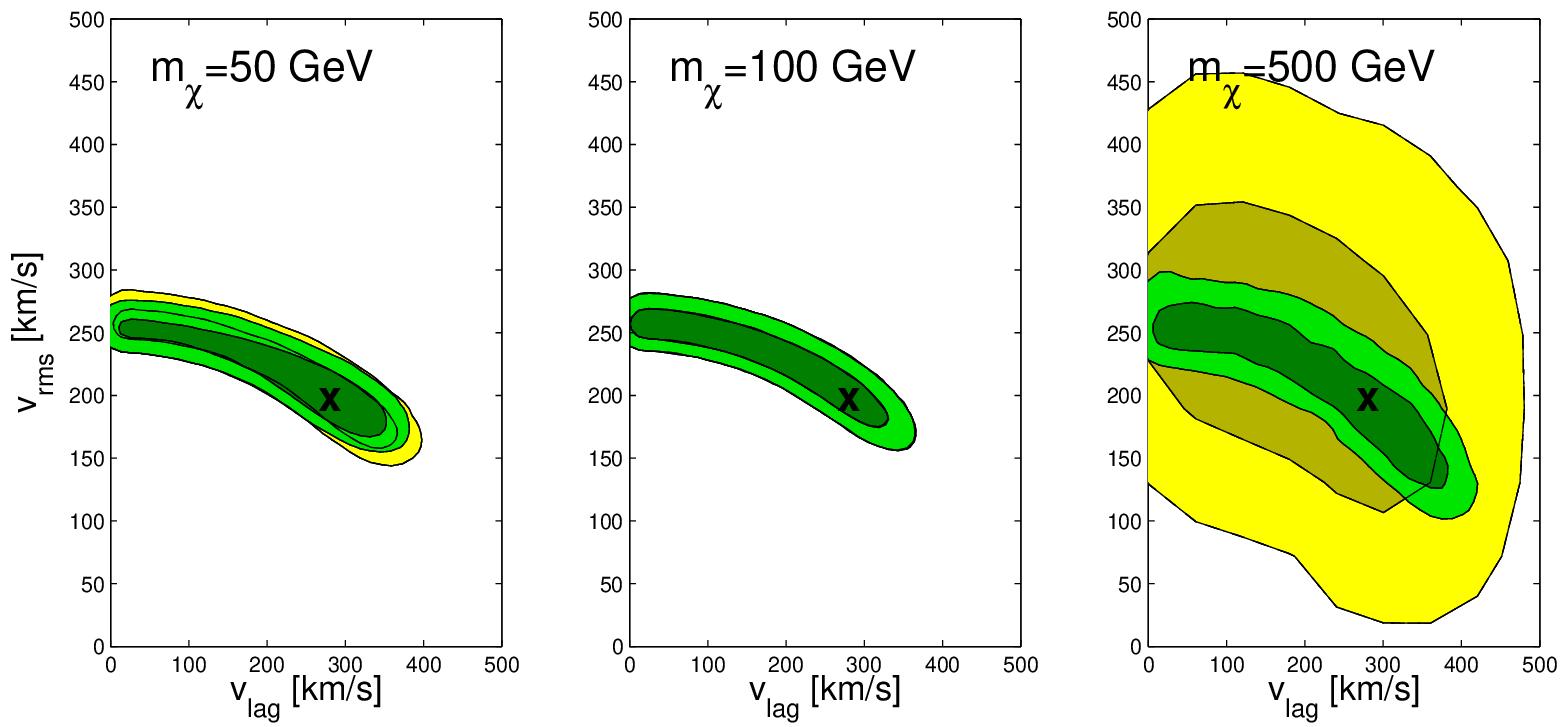}\\
\includegraphics[width=0.61\textwidth]{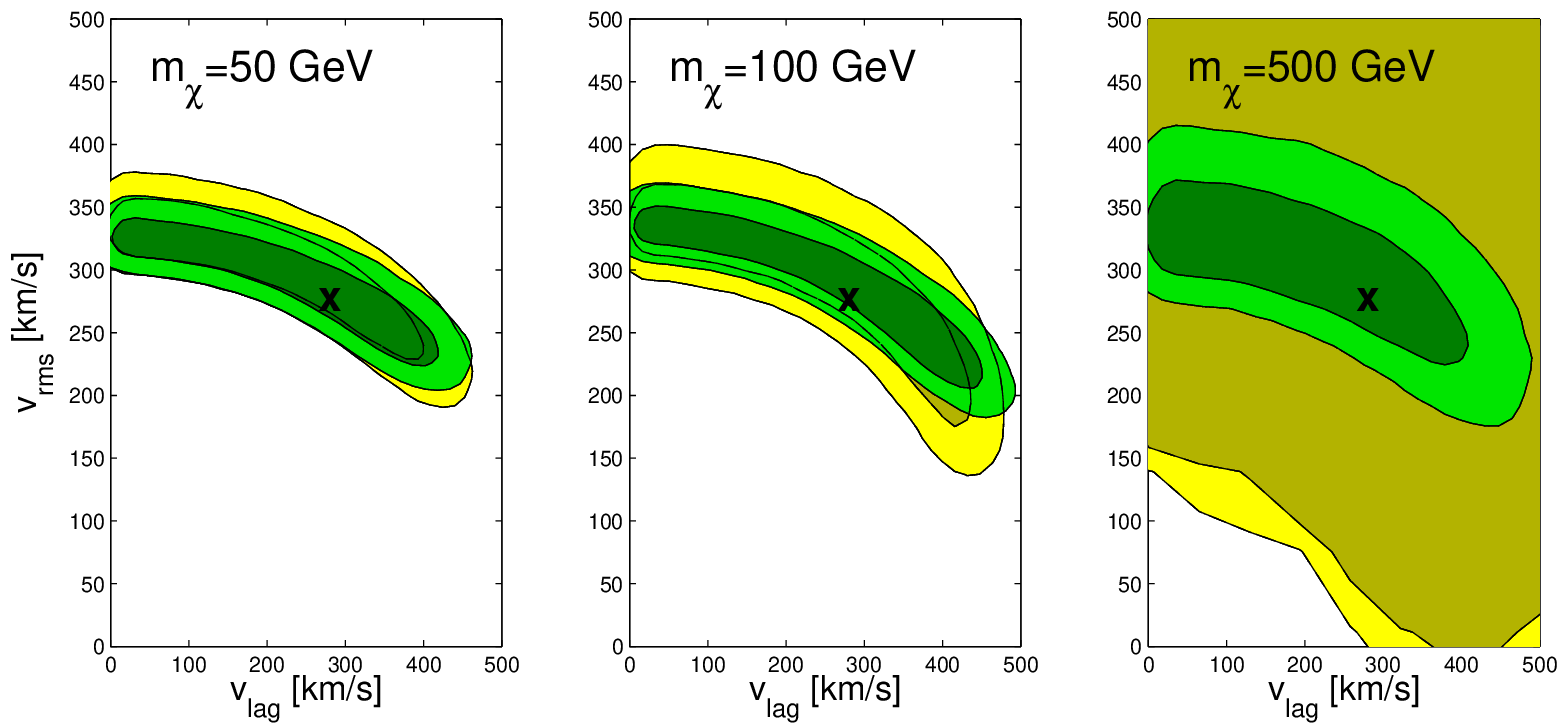}
\caption{\label{fig:oneg_v_prior4}Marginalized probability distributions for \vlag~and \vrms~assuming single Maxwellian speed distributions as both the benchmark models and the hypothesis, and assuming a Gaussian prior on the WIMP mass centered on the benchmark values with the width of the Gaussian equal to $0.1m_\chi$.  The lighter pair of contours is associated with WIMP searches in the analysis windows described in Sec. \ref{subsec:toy}, and the darker pair are associated with extending the analysis window to 1 MeV.  The regions denote 68\% and 95\% C.L. regions.  Each row of figures corresponds to a different WIMP velocity model with (top to bottom): \vlag$=220\hbox{ km s}^{-1}$ \& \vrms$=155\hbox{ km s}^{-1}$; \vlag$=220\hbox{ km s}^{-1}$ \& \vrms$=220\hbox{ km s}^{-1}$; \vlag$=280\hbox{ km s}^{-1}$ \& \vrms$=200\hbox{ km s}^{-1}$; and \vlag$=280\hbox{ km s}^{-1}$ \& \vrms$=280\hbox{ km s}^{-1}$.  Each column represents a different benchmark WIMP mass.}
\end{figure*}

Alternatively, one may view the particle-physics parameters \mwimp~and $D$ as being nuisance parameters if the goal is to determine the WIMP speed distribution as well as possible from the direct-detection data.  In Fig. \ref{fig:oneg_v_prior4}, I consider the marginalized probabilities of \vlag~and \vrms~when imposing a Gaussian prior on \mwimp~centered on the true value, with the width on the prior set to 0.1\mwimp.  This is the range of uncertainty on the WIMP mass one might achieve if supersymmetry is discovered at the Large Hadron Collider (LHC) \cite{baltz2006}.  As in Fig. \ref{fig:oneg_v}, the lighter-filled pair of contours corresponds to the fiducial analysis windows, and the darker-filled contours correspond to setting \Qmax$=1$ MeV.  In general, the mass prior sharpens the \vrms~probability distribution but only alters the constraints in the \vlag-direction a little.   This is somewhat disappointing because it means that we will likely only obtain an upper limit on \vlag~with 2015-era direct-detection experiments.  The mass prior most strongly affects the probability distribution of the speed parameters for high-mass WIMPs because it down weights the speed parameters preferred by the low-\mwimp~tail in the posteriors in Fig. \ref{fig:oneg_v}.  

I have also checked the parameter constraints in the case that the speed distribution deviates significantly from the SHM.  For either a SDD or a high-\vlag, small-\vrms velocity stream, sampling \vrms~and \vlag~linearly in the region $0-2000\hbox{ km s}^{-1}$ in {\sc MultiNest} leads to good constraints on both the WIMP particle-physics parameters and on \vlag~and \vrms.  The only case for which constraints are poor (at least for the fiducial $D$) occurs when the typical recoil energy $Q = \mu_A^2 v_{\mathrm{lag}}^2/m_A$ lies near or below the energy threshold \Qmax~for the lower-threshold experiments.  This constraint improves with larger $D$, though.  Moreover, I have examined the profile likelihoods in addition to the marginalized posteriors, and find the shape of the profile likelihoods and the marginalized posteriors to be broadly consistent regardless of the actual values of \mwimp, \vlag, and \vrms.

\subsection{Multimodal distribution in, Maxwell-Boltzmann distribution out}\label{subsec:2mb}
So far, I have only considered the case in which the hypothesis for the form of speed distribution matches its actual form.  However, there are strong reasons to believe that the DF could be multimodal.  High-resolution dark-matter-only simulations show that there are spatially-varying (on $\sim$ kpc scales) bumps and wiggles in the WIMP velocity distribution, imprints of the halo's accretion history and the tidal stripping of subhalos \cite{vogelsberger2009,kuhlen2010}.  Simulations of Milky Way-mass dark-matter halos that include baryons show that there exists an additional macrostructure, a dark disk formed through the dragging and disruption of satellites in the disk plane of the galaxy \cite{lake1989,read2008}.  Moreover, the Milky Way is still accreting more small halos, which can disrupt and form tidal streams on small scales that have not yet phase mixed.  The key point of this section is to determine how badly \mwimp~and $D$ are biased if one makes the ansatz of a single Maxwell-Boltzmann distribution function even if the distribution function is multimodal.

I examine two multimodal velocity distributions.  First, I consider a model in which half of local WIMPs are described by the SHM and half by the SDD.  I keep $D$ fixed to $3\times 10^{-45}\hbox{ GeV}^{-1}\hbox{cm}^{-1}$, so that the combination of $\rho_\chi$ and \sigmapsi~remain the same as in Sec. \ref{subsec:1mb}.  This model will be called the ``SHM + SDD'' model.  Second, I consider a model in which half the local WIMPs have a SHM distribution function, 30\% have a SDD distribution function, 10\% have a Maxwell-Boltzmann distribution with \vlag$=400\hbox{ km s}^{-1}$ and \vrms$=50\hbox{ km s}^{-1}$, and 10\% have a Maxwell-Boltzmann distribution with \vlag$=500\hbox{ km s}^{-1}$ and \vrms$=50\hbox{ km s}^{-1}$.  These latter two distributions are supposed to represent tidal streams.  This model will be called ``SHM + SDD + 2 streams.''

\begin{figure*}[t]
\centering
\includegraphics[width=0.61\textwidth]{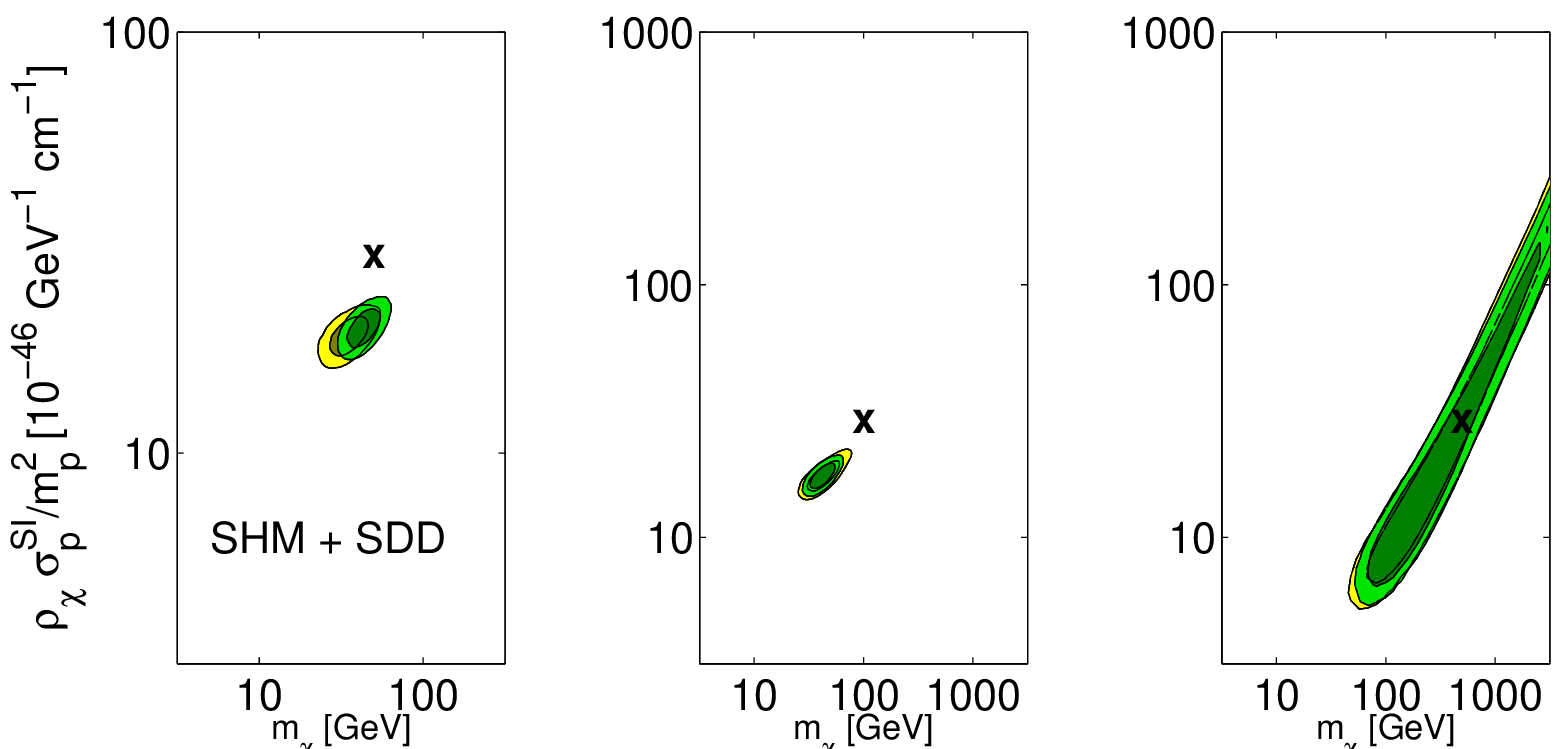}\\
\includegraphics[width=0.61\textwidth]{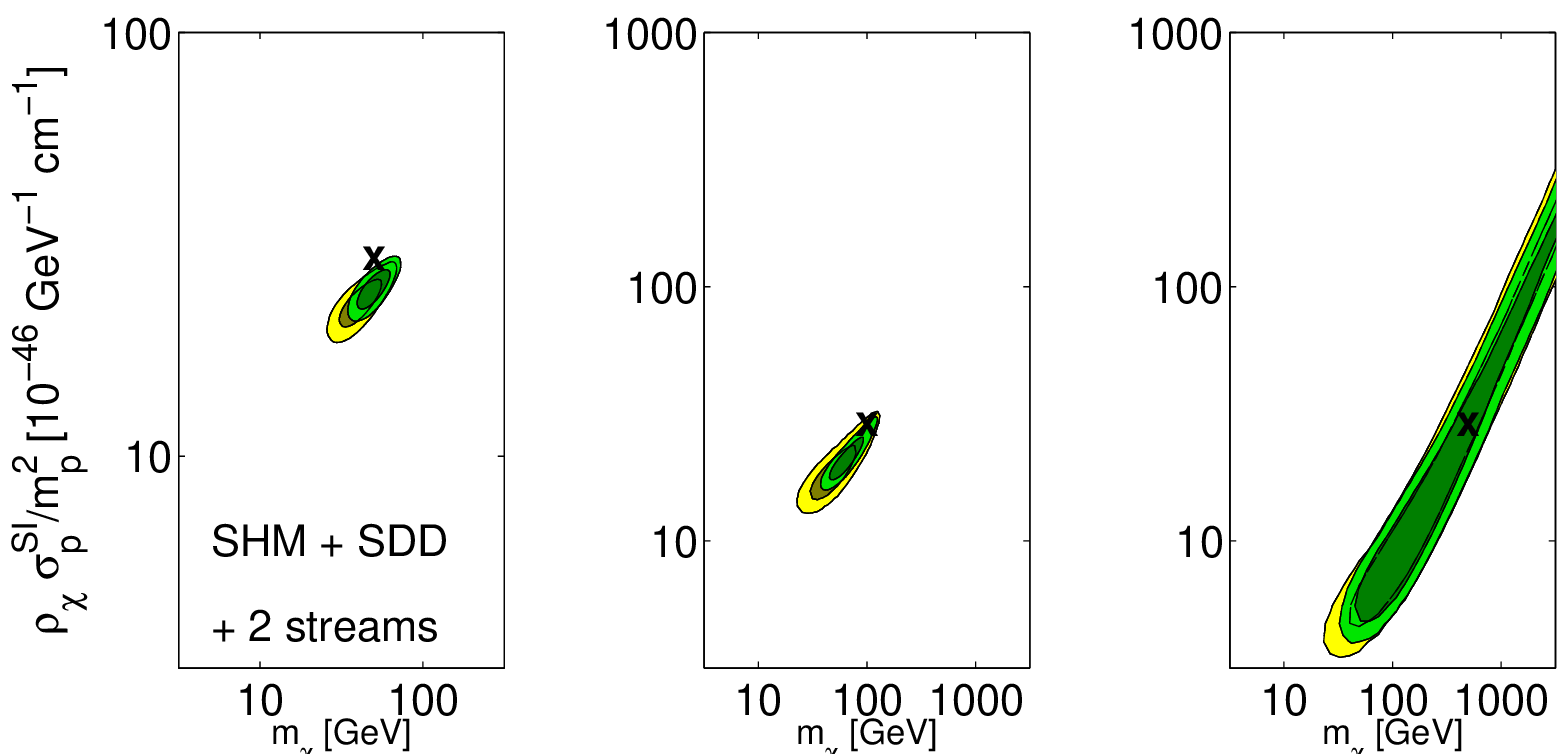}
\caption{\label{fig:oneg_2vel_msigma}Marginalized probability distributions for \mwimp~and $D$.  The 68\% C.L. contours are darker than the 95\% C.L. contours.  The lighter pair of contours is associated with WIMP searches in the analysis windows described in Sec. \ref{subsec:toy}, and the darker pair are associated with extending the analysis window to 1 MeV.  The top panels represent the SHM + SDD benchmark WIMP speed distribution, and the bottom panels represent the SHM + SDD + 2 streams benchmark WIMP speed distribution.  Parameters inferred using the hypothesis of a single Maxwell-Boltzmann population of WIMPs.  See text for details.}
\end{figure*}

I create mock data sets for each of these models for \mwimp$=50,100,$ and 500 GeV, and analyze the data sets with the hypothesis of a single Maxwell-Boltzmann DF.  As in Sec. \ref{subsec:1mb}, I sample \mwimp~and $D$ logarithmically, and \vlag~and \vrms~linearly.  The two-dimensional marginalized probability distributions for \mwimp~and $D$ are shown in Fig. \ref{fig:oneg_2vel_msigma}.  For the SHM + SDD model, the center of the \mwimp$-D$ probability distribution is offset from the true values by $\sim 50\%$ for \mwimp$=50$ and 100 GeV, for either the fiducial \Qmax~or for \Qmax$=1$ MeV.  The direct-detection experiments are generally not overly constraining in the \mwimp$-D$ plane for $m_\chi/m_A \gg 1$, regardless of the true speed distribution.  

The lower panels in Fig. \ref{fig:oneg_2vel_msigma} show the constraints for the SHM + SDD + 2 streams input distribution function.  The centers of the probability contours are offset for \mwimp$=50$ and 100 GeV, although not as much as for the SHM + SDD model.  This is because events from the high-velocity streams populate the high-$Q$ end of the recoil spectrum, which balances out the low-$Q$ dominance of the SDD.

For the two examples explored in this section, using the ansatz of a single, smooth distribution function even if the actual distribution function is multimodal leads to $\sim 50\%$ biases in \mwimp~and $D$.  Although these biases are not as disastrous as they could have been, there are reasons to disfavor using the single-mode distribution-function ansatz for what in reality is likely to be a multimodal distribution function.  First, one would really like to obtain unbiased estimates for the particle-physics parameters for purposes of accurate dark-matter identification.  Additionally, one loses information about the actual speed distribution by forcing a particular form on the data.  This would be a shame, since direct-detection experiments and neutrino-telescope observations of the Sun and Earth provide the \emph{only} way to probe the speed distribution.  This is likely the only way we will ever know if the Milky Way has a dark disk, or if there is a significant amount of microstructure in the Galactic WIMP distribution function.  The next task is to determine if we can find an empirical hypothesis for the local WIMP speed distribution that fits the data better.

\subsection{Empirical speed distributions and hypothesis testing}\label{subsec:vel}
The two goals of this section are to get a sense of how effective empirical speed-distribution models are at recovering the WIMP speed distribution, \mwimp, and $D$; and to determine if one can tell empirically if the DF is not well described by a smooth halo model.  When one does not have an overwhelmingly well-supported theoretical hypothesis, as is the case for the local WIMP DF, it is good to adopt simple, more empirical hypotheses.  This is the approach recommended by the Joint Dark Energy Mission Figure of Merit Science Working Group---instead of forecasting constraints a particular quintessence-inspired equation-of-state evolution function for the dark energy (whose nature is perhaps even less constrained than dark matter), they recommend constraining the equation of states in redshift bins \cite{albrecht2009}.

I adopt a similar approach for the WIMP speed distribution.  In particular, I use a step-function model for the speed distribution in the parameter search.  I focus on constraining the coefficients $g_i$ for a step-function form of the speed distribution,
\begin{eqnarray}\label{eq:stepg}
  \hat{g}(v) = \sum_{i=1}^{N_g} g_i \Theta(v-v_i)\Theta(v_{i+1}-v),
\end{eqnarray}
where $v_i$ is the lower limit of the speed for the $i$th $g(v)$ bin, $v_{i+1}$ is the upper limit.  $\Theta$ is the Heaviside step function.  The hat symbol denotes the fact that this speed distribution is estimated from the data regardless of the true $g(v)$.  In the limit of an infinite number of bins $N_g\rightarrow \infty$, $\hat{g}(v) \rightarrow g(v)$.  In this work, I choose bins of equal size in $v$.  Either the step-function model or the choice of binning may be far from the optimal empirical parametrization of the speed distribution, but these choices for the speed-distribution analysis serve the purpose of providing a good proof of principle for WIMP speed-distribution recovery and model comparison.

For this work, I first consider five bins in speed up to $v=1000\hbox{ km s}^{-1}$.  This upper limit is somewhat larger than the estimated escape speed from the Milky Way in a geocentric frame \cite{smith2007}.  By setting the maximum speed for the speed-distribution bins, I am placing a strong prior that the maximum WIMP speed must lie below that value.  As in the previous sections, I choose the usual benchmarks for WIMP mass, \mwimp$=50,100$, and 500 GeV, and fix $D=3\times 10^{-45}\hbox{ GeV}^{-1}\hbox{cm}^{-1}$.  As before, I sampled those parameters logarithmically using {\sc MultiNest}.  I chose three different benchmarks for the speed distributions for the mock data sets: the SHM, the SHM + SDD, and the SHM + SDD + two high-speed velocity streams (with the same weighting of components as used in Sec. \ref{subsec:2mb}).  I sampled the five velocity-bin coefficients $\{g_i\}$ linearly in the range from 0 to $\{g^{\mathrm{max}}_i\}$, where $g_i^{\mathrm{max}}$ is the maximum value of $g_i$ if all other $g_{j\neq i} = 0$ and satisfying the normalization condition in Eq. (\ref{eq:gvlim}).  While the marginalized 68\% and 95\% confidence-level regions in the speed coefficients are not dramatically different if one samples the $\{g_i\}$ logarithmically, the marginalized contours generally follow the shape of the profile likelihood better for linear scans in $\{g_i\}$.

\begin{figure*}[t]
\centering
\includegraphics[width=0.8\textwidth]{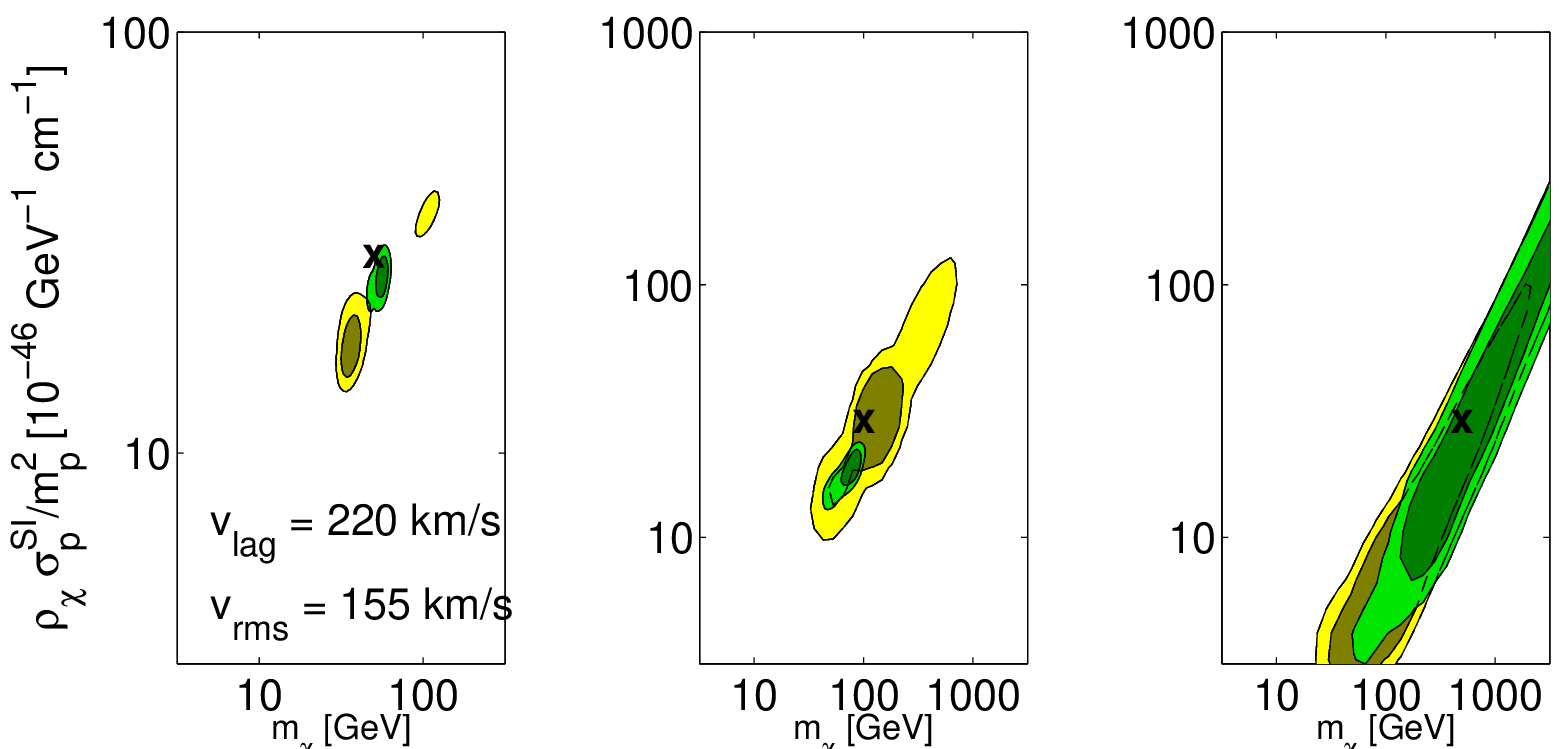}
\caption{\label{fig:vel_msigma}Marginalized probability distributions for \mwimp~and $D$~with the SHM as the velocity benchmark and analyzed with the five-bin step-function speed-distribution hypothesis.  Each panel represents a different benchmark \mwimp.  The lighter pair of contours represents 68\% and 95\% C.L. regions based on the analysis windows described in Sec. \ref{subsec:toy}, and the darker pair of contours are the results if the analysis windows are extended to 1 MeV.}
\end{figure*}

\begin{figure}[t]
\centering
\includegraphics[width=0.47\textwidth]{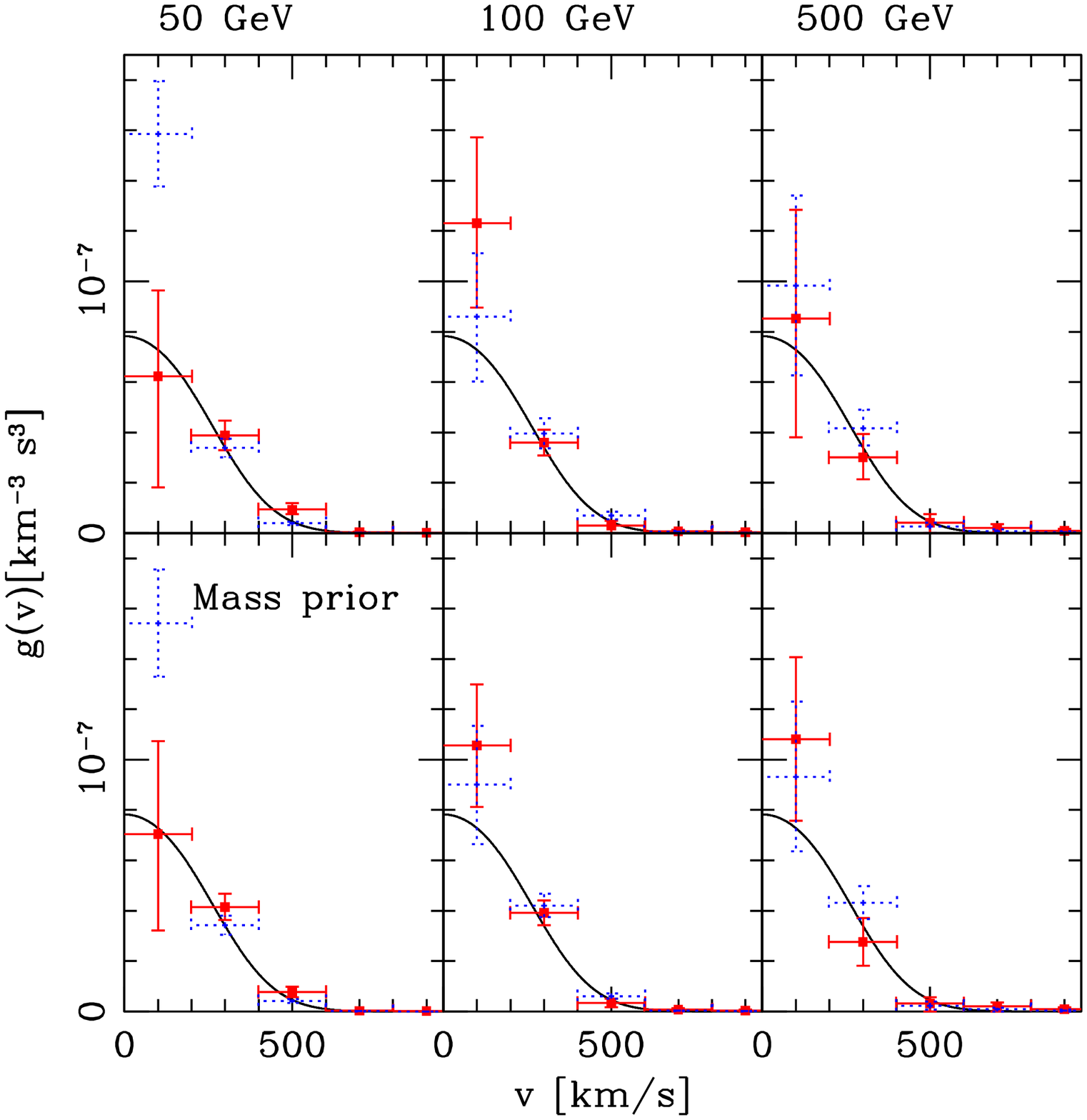}
\caption{\label{fig:vel_v}Inferred WIMP speed distributions for the SHM benchmark model.  The solid error bars denote the 68\% C.L. region for each $g_i$ using the fiducial analysis windows, and the dashed error bars show the same but for $Q_{\mathrm{max}} = 1$ MeV.  The upper panels show the speed constraints when the WIMP mass and cross section are sampled logarithmically, and the bottom panels show the speed constraints when there is an additional 10\% Gaussian prior on the WIMP mass.  The solid line denotes the benchmark speed distribution.}
\end{figure}

The first benchmark speed distribution I consider is the SHM.  The constraints in the \mwimp$-D$ plane are shown in Fig. \ref{fig:vel_msigma}, and the constraints on $\{g_i\}$ are shown in Fig. \ref{fig:vel_v}.  In Fig. \ref{fig:vel_msigma}, I show the marginalized probabilities for the fiducial \Qmax~with the set of light-colored filled regions, and the marginalized probabilities for \Qmax$=1$ MeV with the darker pair of regions.  The first thing to note is that the parameter uncertainties are no larger than those found in Sec. \ref{subsec:1mb}, although they are biased in the cases of \mwimp$=50$ and 100 GeV.  The bias decreases with increasing \Qmax, though.  The second thing to note is that the shape of the degeneracy contours is quite different than with the Maxwell-Boltzmann ansatz used in Sec. \ref{subsec:1mb}.  This is because the shape of the mapping between \mwimp~and $D$ and a fixed recoil spectrum depends on the form of the speed distribution.  Third, for $m_\chi = 50$ GeV there are disconnected regions.  This is an artifact of the ``realization noise'' in the data.

The reconstructed speed distributions are shown in Fig. \ref{fig:vel_v}.  Each column in the figure represents a different WIMP mass.  The error bars represent the marginalized 68\% probability limits for each $g_i$.  Note that the probability contours are in fact correlated.  The solid error bars denote the limits obtained with the fiducial \Qmax, and the dotted error bars denote those obtained if \Qmax$=1$ MeV.  In the upper panels, the WIMP mass is only constrained to be somewhere between 1 MeV and 100 TeV, but in the bottom panels, I impose a Gaussian prior on the WIMP mass centered on the true value and with a width of $0.1$\mwimp.  The solid line shows the SHM speed distribution.  In general, using the higher \Qmax~leads to better fits to the SHM speed distribution, with the exception of the case in which \mwimp$=50$ GeV.  I note that a similar trend towards a larger low-speed population is also seen in the Maxwell-Boltzmann analysis in Sec. \ref{subsec:1mb} for this particular benchmark.  Figures \ref{fig:oneg_v} and \ref{fig:oneg_v_prior4} show that the true speed-parameter point barely lies within the 95\% C.L. contour.  This high inferred density of low-speed particles is an artifact of this particular realization of the data for this set of benchmark parameters.

Although the inferred speed distributions look reasonable, one might want to ask if the inferred speed distribution were consistent with Maxwell-Boltzmann distribution.  The issue of model selection is tricky for both the frequentist and Bayesian perspectives if one cannot use $\chi^2$ to determine the goodness of fit (e.g., \cite{trotta2008,brewer2009}).  In general, the goal is to determine the \emph{relative} fit between hypotheses instead of determining the absolute quality of fit for a single hypothesis.  I use three different criteria to assess the relative quality of fit between the single Maxwell-Boltzmann and step-function speed-distribution hypotheses: the Bayes factor, the Akaike information criterion (AIC) \cite{akaike1974}, and the Bayes information criterion (BIC) \cite{schwarz1978}.  However, for reasons stated below, I will emphasize the Bayes factor in particular.

In the Bayesian context, the ratio of Bayesian evidences [Eq. (\ref{eq:evidence})] for two hypotheses (``Bayes factor'') is often used to determine if one hypothesis fits the data better than the other.  Since the Bayesian evidence is just the average likelihood over the parameter space (weighted by the prior), the better-fit hypothesis is assumed to be the one with the higher average likelihood, regardless of maximum likelihood $\mathcal{L}_{\mathrm{max}}$.  This means that models with fewer parameters are generally preferred (the ``Occam's razor'' hypothesis---simpler models are better).  Technically, the Bayes factor is not strictly the ratio of evidences, but is the ratio of evidences multiplied by the ratio of the priors on the hypotheses.  Quantifying the belief in the hypotheses is something I will not get into in this work, but Ref. \cite{brewer2009} provides an interesting introduction to the subject.  For now, I will assume that the hypotheses are equally probable.

In general, the evidence is prior and parameter dependent.  In the present case, determining whether the step-function hypothesis fits better than a Maxwell-Boltzmann distribution, the fact that WIMPs cannot travel with infinite speed allows one to at least define a reasonable parameter volume.  For the step-function speed model, the $\{g_i\}$ cannot exceed $\{g_i^{\mathrm{max}}\}$.  This provides a natural volume to use, and was the volume I used to for the parameter search.  As with the parameter search, I use flat priors on $\{g_i\}$ to calculate the evidence.  To calculate the evidence for the Maxwell-Boltzmann model, I use the same priors and parameter-space volume as used in the parameter search in Sec. \ref{subsec:1mb}.  The upper bound for \vlag~and \vrms~are well above the escape speed from the Galaxy.

Although I use the Bayes factor
\begin{eqnarray}
  B = \frac{\mathcal{Z}(1MB)}{\mathcal{Z}(\{g_i\})}
\end{eqnarray}
to get a sense the relative fit of the single Maxwell-Boltzmann (1MB) and step-function ($\{g_i\}$) models, more in-depth studies are necessary to determine if this is really the best fit criterion for direct-detection data.  Moreover, even though the way in which I have defined the prior volume is reasonable, it may not be the best; \vlag~and \vrms~are a completely different way of parametrizing a speed distribution than $\{g_i\}$.  However, as I show below, the Bayes factor seems to be a not unreasonable criterion by which to classify fits.

Second, I consider the AIC, which is approximated as
\begin{eqnarray}
  \mathrm{AIC} = -2\ln\mathcal{L}_{\mathrm{max}} + 2 N_p,
\end{eqnarray}
where $\mathcal{L}_{\mathrm{max}}$ is the maximum likelihood for the data given the hypothesis, and $N_p$ is the number of parameters of the hypothesis.  The AIC is meant to minimize the Kullback-Leibler information entropy \cite{kullback1951}, and so the hypothesis with the smallest AIC is preferred.  As with most Bayesian model-selection criteria, the AIC penalizes the introduction of additional parameters, but not as much as the BIC, the third model-selection criterion I consider, which is defined as
\begin{eqnarray}
  \mathrm{BIC} = -2\ln\mathcal{L}_{\mathrm{max}} + N_p \ln(N_o).
\end{eqnarray}
Here, $N_o$ is the observed number of events.  For the data sets I consider, there are between $\sim 200$ and $\sim 700$ total events, which gives $\ln(N_o) \sim 6$.  In the limit that the posterior is a multivariate Gaussian and that the data are independent and identically distributed, the Bayesian evidence and BIC are equivalent in terms of describing the quality of the fit \cite{trotta2008}.

Even though I consider all three Bayesian model-selection criteria below, I emphasize the Bayes factor because it is easiest to interpret and most likely to select the better model.  The AIC does not necessarily select the correct model even if one had an infinite, unbiased data set \cite{kashyap1980}.  The issue with the BIC is that the posteriors for the direct-detection data sets are clearly not well described by multivariate Gaussians, and so it is not clear how then to interpret the BIC.  In cosmology, the Bayes factor is the preferred Bayesian model-selection criterion \cite{liddle2006,liddle2007,trotta2007}.  I show the Bayes factor for each benchmark model in Table \ref{tab:bayes}.


\begin{table}[t]
\caption{\label{tab:bayes}Bayes' factor for benchmark speed distributions and WIMP masses}
\begin{ruledtabular}

\begin{tabular}{lrrr}

  Benchmark speed distributions &$m_\chi$ [GeV] &$Q_{\mathrm{max}}$ &$\ln B$ \\
\hline
SHM &50 &fiducial &2.7 \\
   &50 &1 MeV &2.1 \\
   &100 &fiducial &0.8\\
   &100 &1 MeV &2.8 \\
   &500 &fiducial &-2.5\\
   &500 &1 MeV &-2.1 \\
\hline
SHM + SDD &50 &fiducial &-3.1\\
   &50 &1 MeV &-6.6 \\
   &100 &fiducial &-4.3\\
   &100 &1 MeV &-6.3\\
   &500 &fiducial &-1.8\\
   &500 &1 MeV &-3.1\\  
\hline
SHM + SDD + 2 streams &50 &fiducial &-2.9\\
   &50 &1 MeV & -7.4\\
   &100 &fiducial &-1.9\\
   &100 &1 MeV &-8.5 \\
   &500 &fiducial &-3.9\\
   &500 &1 MeV &-3.4 \\

\end{tabular}

\end{ruledtabular}
\end{table}

For all the SHM data sets, the SHM is preferred over the five-bin step-function speed-distribution hypothesis by the AIC and the BIC.  However, the preference is not especially strong according to the Bayes factor.  The mock data sets with the highest $B$ are those with \mwimp$=50$ GeV with the fiducial \Qmax~and \mwimp$=100$ GeV with \Qmax$=1$ MeV, for which $\ln(B) \approx 3$, which is almost considered ``moderate'' evidence on the Jeffreys' scale in favor of the SHM \cite{trotta2008}.  In the cases in which $m_\chi = 500$ GeV, there is weak evidence for the step-function model; it is, however, not especially significant since any $|\ln B| < 3$ is considered weak evidence.

\begin{figure*}[t]
\centering
\includegraphics[width=0.8\textwidth]{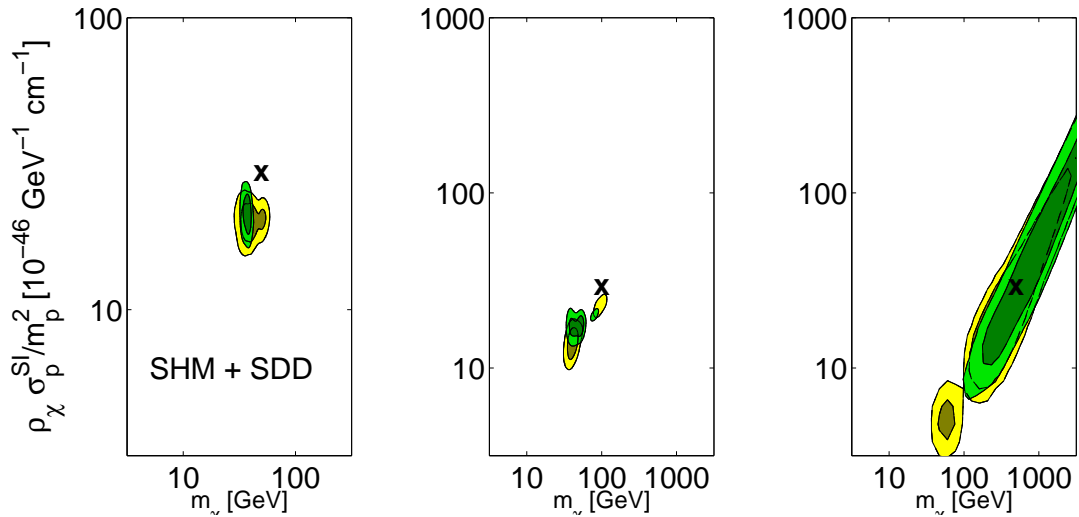}
\caption{\label{fig:vel_2vel_msigma}Marginalized probability distributions for \mwimp~and $D$~for the SHM + SDD benchmark speed-distribution model and the five-bin step-function speed-distribution hypothesis.  Each panel represents a different benchmark \mwimp.  The lighter pair of contours represents 68\% and 95\% C.L. regions based on the analysis windows described in Sec. \ref{subsec:toy}, and the darker pair of contours are the results if the analysis windows are extended to 1 MeV.}
\end{figure*}

\begin{figure}[t]
\centering
\includegraphics[width=0.47\textwidth]{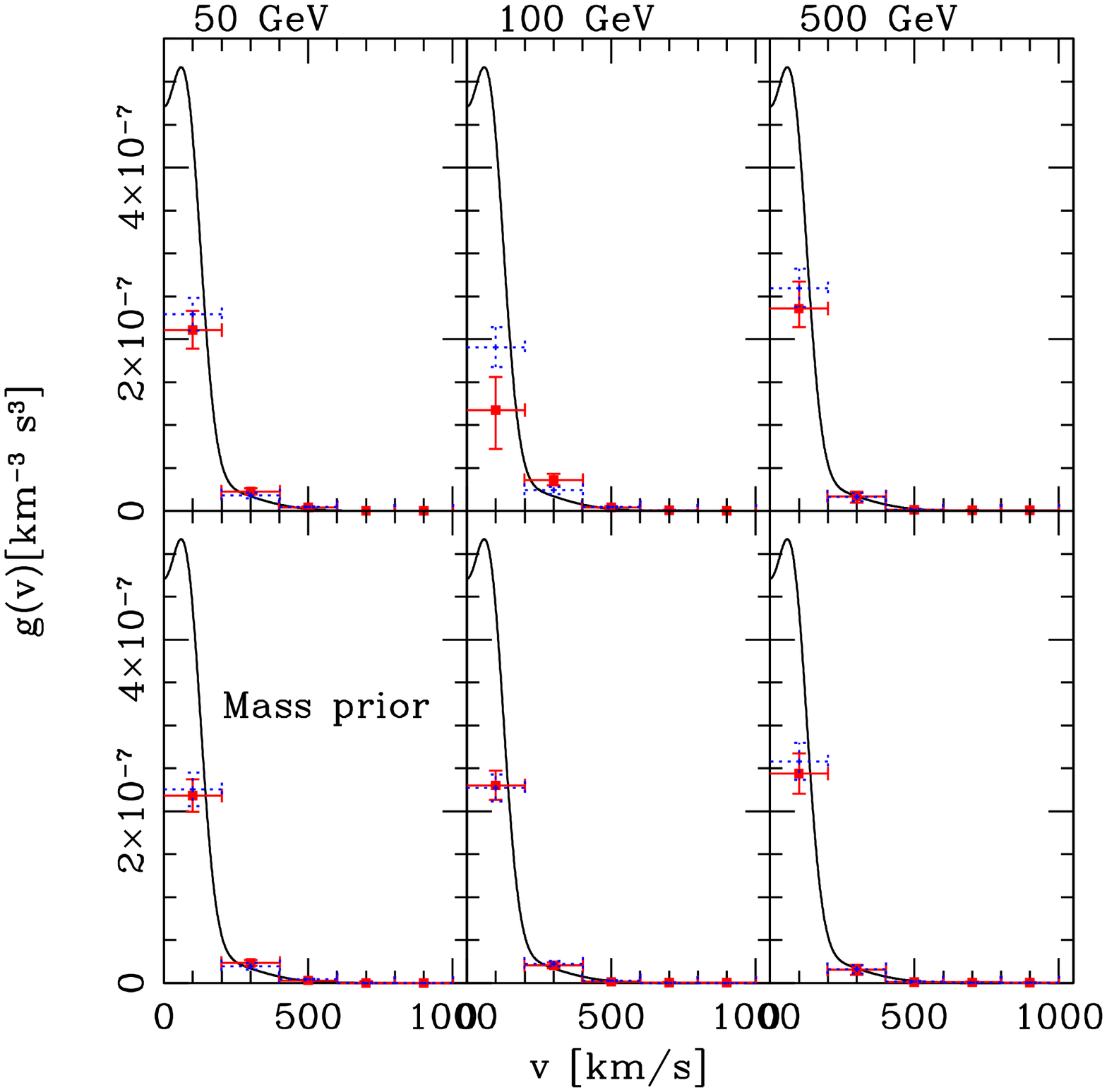}
\caption{\label{fig:vel_2vel_v}Inferred WIMP speed distributions for the SHM + SDD benchmark model.  The solid error bars denote the 68\% C.L. region for each $g_i$ using the fiducial analysis windows, and the dashed error bars show the same but for $Q_{\mathrm{max}} = 1$ MeV.  The upper panels show the speed constraints when the WIMP mass and cross section are sampled logarithmically, and the bottom panels show the speed constraints when there is an additional 10\% Gaussian prior on the WIMP mass.  The solid line denotes the benchmark speed distribution.}
\end{figure}

Next, I consider the multimodal distribution functions I explored in Sec. \ref{subsec:2mb}.  The constraints for the SHM + SDD benchmark model in the \mwimp$-D$ plane are shown in Fig. \ref{fig:vel_2vel_msigma}, and the speed-distribution fits are shown in Fig. \ref{fig:vel_2vel_v}.  As in Fig. \ref{fig:vel_msigma}, the lighter pair of contours indicate the marginalized probabilities of the parameters for the fiducial values of \Qmax, and the darker pair correspond to setting \Qmax$=1$ MeV.  The probability contours in the \mwimp$-D$ plane are offset from the true point, with the exception of the \mwimp$=500$ GeV cases.  The offsets are somewhat less than if one were to apply the ansatz that the velocity distribution is Maxwell-Boltzmann (Fig. \ref{fig:oneg_2vel_msigma}), but not much.  The offsets are lower if one uses a higher \Qmax.  The probability contours have a different shape than those resulting from the hypothesis that the velocities have a Maxwell-Boltzmann distribution.

The inferred speed distribution is shown in Fig. \ref{fig:vel_2vel_v}, in which the lines and error bars have the same meaning as in Fig. \ref{fig:vel_v}.  While the inferred speed distribution is reasonable for the speed bins above 200 $\hbox{km s}^{-1}$, and although the fits appear better for the cases of high \Qmax, the first speed bin is always systematically low.  The main reason for this is that the speeds below approximately 100$\hbox{ km s}^{-1}$ are actually quite poorly constrained by the data.  This is because the typical recoil energy for a WIMP moving 100$\hbox{ km s}^{-1}$ with respect to the experiment is
\begin{eqnarray}
Q\sim 0.1 \frac{\mu_A}{\hbox{1 GeV}}\frac{\mu_A}{m_A}\hbox{ keV}.
\end{eqnarray}
Thus, most of the low-speed WIMPs will scatter below \Qmin, especially for the argon experiment, and so the lowest-speed bin in Fig. \ref{fig:vel_2vel_v} actually reflects WIMPs in the speed range $v=100-200\hbox{ km s}^{-1}$.

\begin{figure}[t]
\centering
\includegraphics[width=0.47\textwidth]{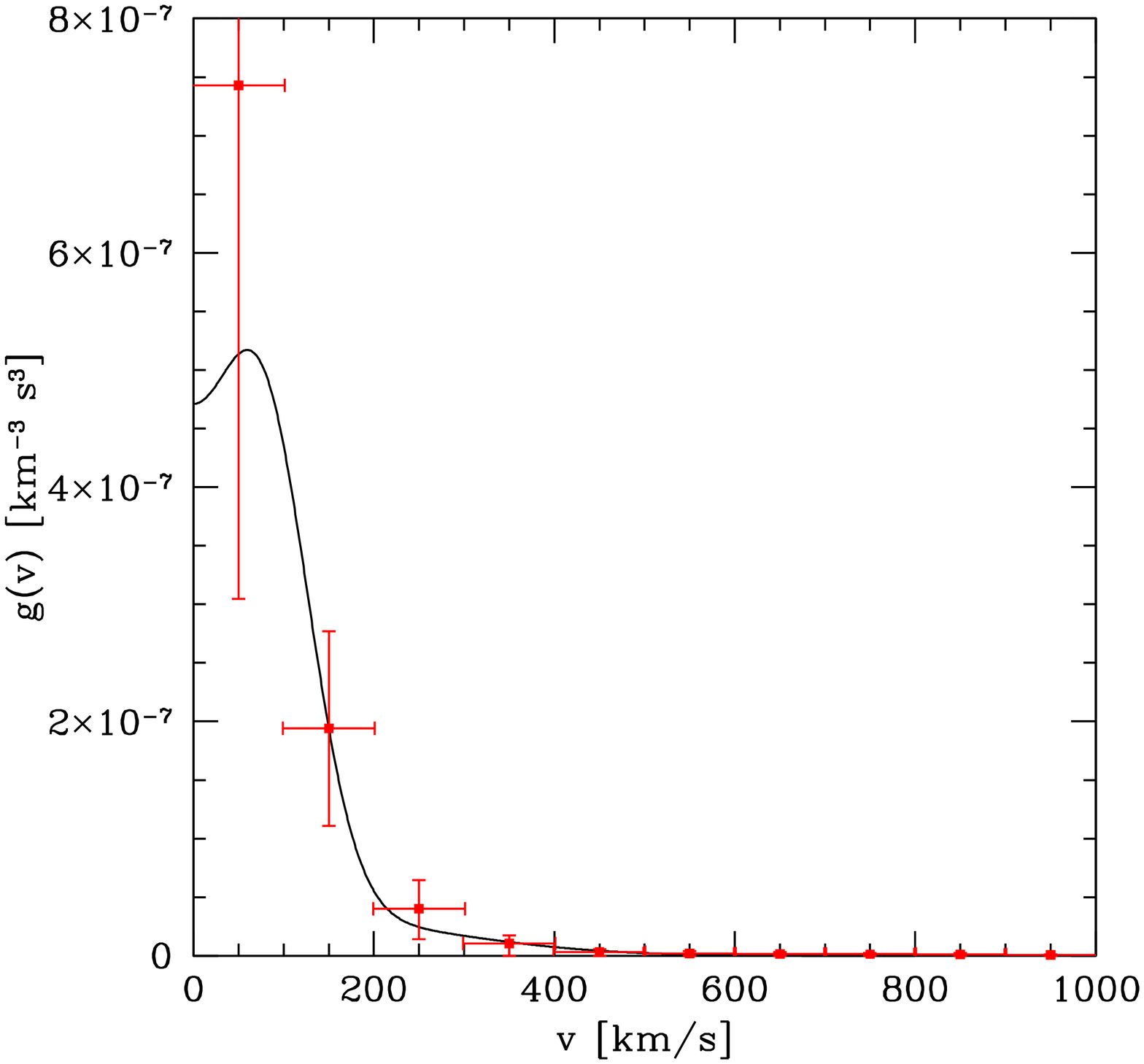}
\caption{\label{fig:vel_10bins}Speed distribution inferred from mock data for the SHM + SDD model and \mwimp$=500$ GeV.  There are ten bins equally sized in $v$ up to $v=1000\hbox{ km s}^{-1}$.}
\end{figure}

This point is further illustrated in Fig. \ref{fig:vel_10bins}, in which I show the inferred speed distribution for a step-function speed-distribution hypothesis with ten speed bins.  In this case, the true WIMP mass is 500 GeV and I used the fiducial \Qmax~for the experiments.  In this figure, the 68\% C.L. region for the lowest-speed bin is enormous compared to the other bins.  The $v=100-200\hbox{ km s}^{-1}$ bin is well centered on its true value.  Thus, the systematically low value of the speed distribution for the lowest-speed bin in the five-bin model is an artifact of the fact that the experiments cannot really constrain WIMPs that scatter below threshold.

The lack of sensitivity to the lowest-speed WIMPs also explains several features of the $m_\chi - D$ parameter constraints.  For the $m_\chi = 50$ and 100 GeV cases, both \mwimp~and $D$ tend to be too low, with multiple peaks in the posterior for $m_\chi = 100$ and 500 GeV in Fig. \ref{fig:vel_2vel_msigma} (and for $m_\chi = 50$ GeV in Fig. \ref{fig:vel_msigma}).  For the SuperCDMS- and LUX-inspired experiments, the threshold recoil energy \Qmax~lies right on the transition between the lowest two speed bins for $m_\chi \approx 80$ GeV in the five-bin hypothesis.  For lower-mass WIMPs, those experiments are completely insensitive to the lowest-speed bin, and for higher mass WIMPs, the experiments are sensitive to speeds only at the upper edge of the lowest-speed bin.  However, for the SHM + SDD model, a relatively large fraction of the WIMPs are actually in the lowest-speed bin.  The multiple peaks in the posterior appear to be associated with this transition in sensitivity to the lowest-speed bin, especially since this particular empirical description of the speed distribution is discontinuous.  The cross section is biased low for the following reason.  Since the differential event rate is highest near threshold, the constraint on the speed bin just above threshold is strong and is more influenced by the lower speed WIMPs in the bin.  Since the event rate goes as $\sim \int g(v) v dv$ and the number density of particles goes as $\sim g(v) v^2 dv$, the number of WIMPs in the second-lowest-speed bin is biased high while the number of WIMPs in the lowest-speed bin is biased low (also due to the fact that the experiments are sensitive only to speeds at the upper edge of the speed bin).  The cross section must drop to compensate for the relatively high number of WIMPs inferred in the second-lowest-speed bin.

Next, I consider the question of model selection.  I calculate the Bayes factor, AIC, and BIC for the SHM + SDD data sets.  I find that the Bayes factor indicates that the step-function model is a better fit for each of the six ensembles of mock data sets.  The Bayes factor is most significant for \mwimp$=50$ and 100 GeV for \Qmax$=1$ MeV.  In those cases, $\ln(B) = -7$ to $-6$ (Table \ref{tab:bayes}), which is considered ``strong evidence'' on the Jeffreys' scale \cite{trotta2008}.  All ensembles of data sets with the exception of the single ensemble with \mwimp$=500$ GeV and the fiducial \Qmax~indicate a lower AIC for the step-function model than the Maxwell-Boltzmann distribution.  However, only those mock data sets corresponding to \mwimp=$50$ or 100 GeV with \Qmax$=1$ MeV additionally have a lower BIC for the step-function model than for the Maxwell-Boltzmann distribution.  Thus, while the Bayes factor and AIC generally show that the Maxwell-Boltzmann distribution is disfavored for this particular two-component velocity distribution, it is generally only moderately disfavored relative to the step-function model unless \Qmax~is large.

\begin{figure*}[t]
\centering
\includegraphics[width=0.8\textwidth]{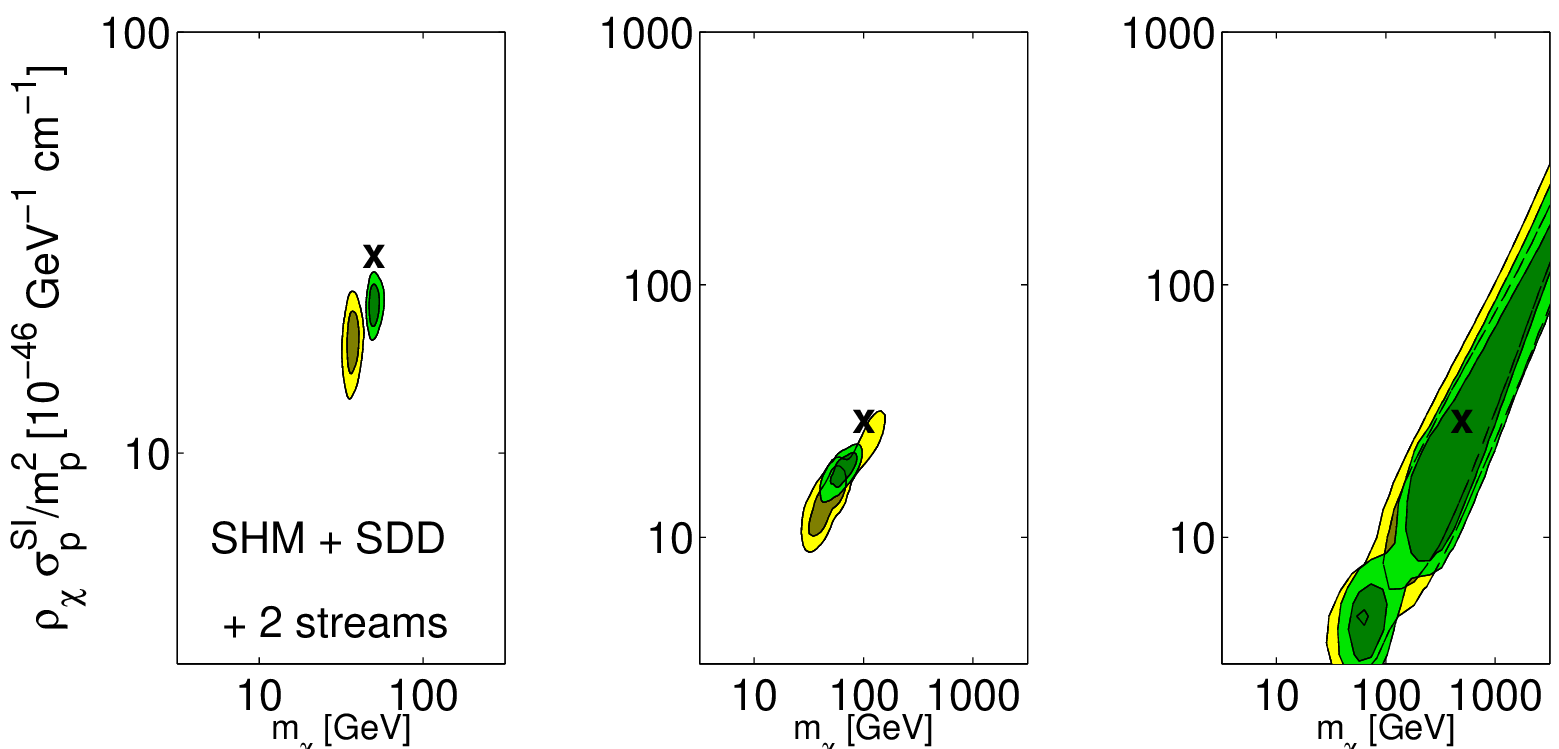}
\caption{\label{fig:vel_4vel_msigma}Marginalized probability distributions for \mwimp~and $D$~for the SHM + SDD + 2 streams benchmark model and the five-bin step-function speed-distribution hypothesis.  Each panel represents a different input mass.  The lighter pair of contours represents 68\% and 95\% C.L. regions based on the analysis windows described in Sec. \ref{subsec:toy}, and the darker pair of contours are the results if the analysis windows are extended to 1 MeV.}
\end{figure*}

\begin{figure}[t]
\centering
\includegraphics[width=0.47\textwidth]{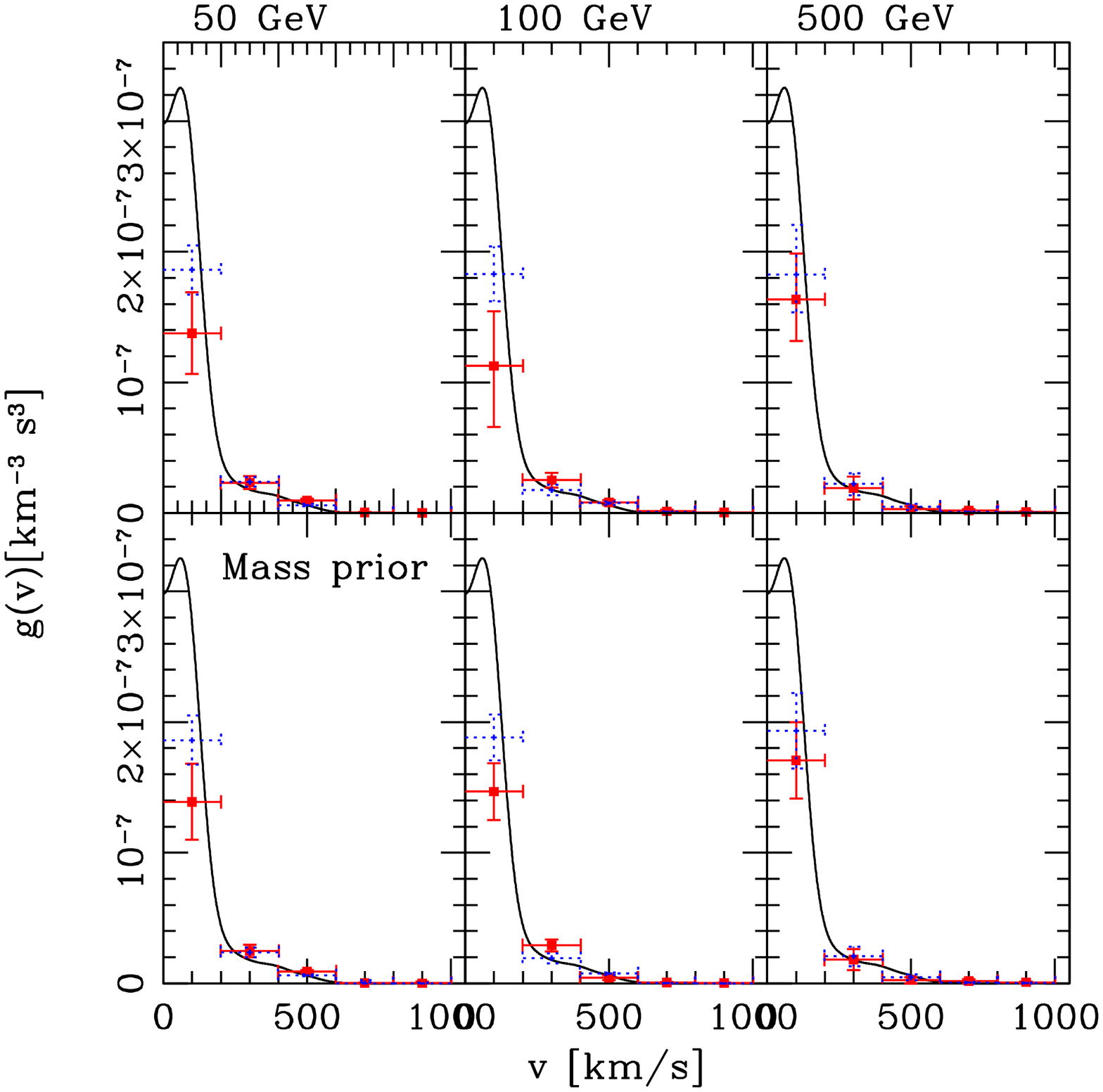}
\caption{\label{fig:vel_4vel_v}Inferred WIMP speed distributions for the SHM + SDD + 2 streams benchmark model.  The solid error bars denote the 68\% C.L. region for each $g_i$ using the fiducial analysis windows, and the dashed error bars show the same but for $Q_{\mathrm{max}} = 1$ MeV.  The upper panels show the speed constraints when the WIMP mass and cross section are sampled logarithmically, and the bottom panels show the speed constraints when there is an additional 10\% Gaussian prior on the WIMP mass.  The solid line denotes the benchmark speed distribution.}
\end{figure}

Last, I consider the SHM + SDD + 2 streams model.  The probability contours in the \mwimp$-D$ plane for several values of \mwimp~are shown in Fig. \ref{fig:vel_4vel_msigma}, and the speed distribution is shown in Fig. \ref{fig:vel_4vel_v}.  As for the SHM + SDD case, the contours in the \mwimp$-D$ plane are typically slightly offset from the true point in parameter space, but are less offset for higher \Qmax.  Also as for the SHM + DD case, the lowest-speed bin in is systematically low due to the poor constraints on the lowest-speed WIMPs, although the higher-speed bins are well centered on the true speed distribution.

The model-selection patterns also follow those of the SHM + SDD benchmark speed distribution.  There is only one exception to the patterns of the SHM + SDD findings.  The case of \mwimp$=100$ GeV with fiducial \Qmax~has AIC and BIC that prefer the Maxwell-Boltzmann velocity distribution model (although in the case of the AIC, the difference between the two models is quite small, $<2$), and the Bayes factor is $\ln(B) = -1.9$, which indicates a weak-to-moderate preference for the step-function model.  Otherwise, the trends in $\ln(B)$, AIC, and BIC for the SHM + SDD input velocity model hold for this more complicated input velocity model, too.  The Bayes' factors for all benchmarks are given in Table \ref{tab:bayes}.

There are a few final points I would like to address in this section.  First, although I have shown for the mock data sets with multimodal velocity distributions that one may reasonably reconstruct a speed distribution (taking care with the low-speed end for which the experiments have little constraining power) using the step-function hypothesis, there is still the issue that the \mwimp$-D$ probability distribution is offset from the true value.  In fact, the true point lies outside of the 95\% C.L. contour in most cases I considered.  This is because, even though the five-bin step-function speed-distribution hypothesis is a better fit to the data, it is by no means the best fit speed-distribution hypothesis for the data.  In fact, the discontinuous nature of the speed-distribution hypothesis is clearly not physical and is responsible for some of the odder features of the probability contours, as discussed in this section.

\begin{figure}
  \centering
  \includegraphics[width=0.42\textwidth]{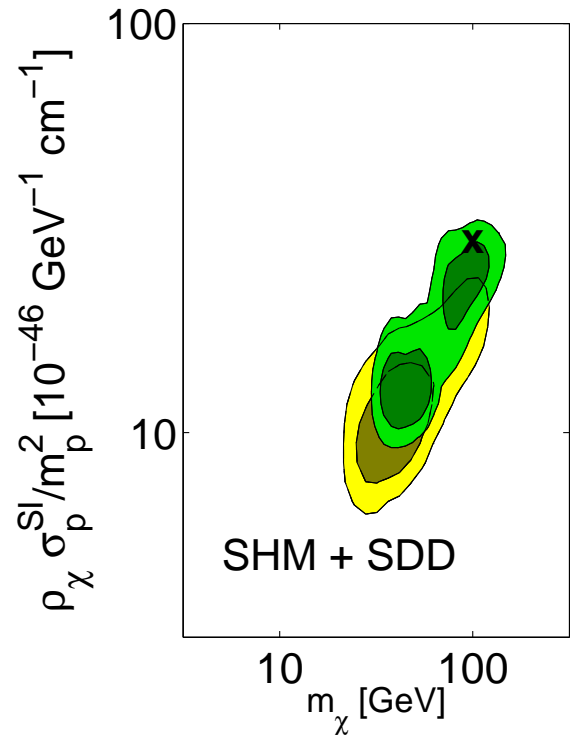}
  \caption{\label{fig:10bin}Marginalized probability distributions for \mwimp~and $D$~for $m_\chi = 100$ GeV and SHM + SDD benchmark model and the ten-bin step-function speed-distribution hypothesis.  The lighter pair of contours represents 68\% and 95\% C.L. regions based on the analysis windows described in Sec. \ref{subsec:toy}, and the darker pair of contours are the results if the analysis windows are extended to 1 MeV.}
\end{figure}

One may ask if one does better with a larger number of speed bins.  Even though the speed-distribution hypothesis is still discontinuous, it is a better representation of a continuous function.  I ran a set of tests in which I doubled the number of speed bins, from five to ten.  The maximum likelihood $\mathcal{L}_{\mathrm{max}}$ barely improved between the two sets of analyses ($|\Delta \ln \mathcal{L}_{\mathrm{max}}| \lesssim 3$), meaning that the AIC and BIC model-selection criteria would prefer the five-bin model over the ten-bin model.  The only case in which $\mathcal{L}_{\mathrm{max}}$ increased enough that the AIC preferred the 10-bin model was for the SHM + SDD benchmark with $m_\chi = 100$ GeV and \Qmax$=1$ MeV.  The Bayes factor
\begin{eqnarray}
  B_{\mathrm{bin}} = \frac{\mathcal{Z}(\hbox{5 bins})}{\mathcal{Z}(\hbox{10 bins})}
\end{eqnarray}
was ranged from nearly 1 (no preference in favor of either model) to $\ln(B_{\mathrm{bin}}) = 3$, which indicates moderate preference for the five-bin model.  Even for the one case in which the AIC indicated the ten-bin model was a better fit, the Bayes factor indicated a preference for the five-bin model.  Usually, the Bayes factor was less significant in distinguishing between the hypotheses of the number of bins in the step-function speed-distribution model than distinguishing between the five-bin model and the Maxwell-Boltzmann distribution.  It appears that for the mock data sets I considered (and what is likely to hold true for the first years of real data), ten speed bins is likely overkill.  The one interesting feature of the ten-bin model was that the \mwimp$-D$ probability distributions were better centered on the true value than for the five-bin model, although the posterior is still multimodal.  This is illustrated in Fig. \ref{fig:10bin}, in which I show the probability distributions for the SHM + SDD benchmark for $m_\chi = 100$ GeV.  However, this needs to be explored for more ensembles of mock data sets to see if that is generally true. 

Although I have shown that one may achieve demonstrably better fits to multimodal velocity distributions using the step-function speed-distribution hypothesis, I have not shown that this is truly the best empirical speed-distribution model one could use.  In fact, the discontinuous nature of this empirical function has clear downsides, in addition to only being able to approximate theoretically-inspired functional forms for the speed distribution in the limit of many bins.  In practice, it is likely that direct-detection data sets will need to be analyzed with a variety of empirical hypotheses for the speed distribution in order to achieve the best, unbiased constraints on both the particle-physics parameters (\mwimp, $D$) and the speed distribution.  I leave the development of a strategy for optimal model selection to future work.

The key points of this section are that one may obtain a reasonable estimate of the WIMP speed distribution using these simple step-function speed-distribution models, and that one may distinguish between speed-distribution models using Bayesian model-selection criteria.  In particular, the simple step-function model is moderately to strongly preferred over the single-velocity-component model for the benchmark multimodal velocity distributions.

\section{Discussion}\label{sec:discussion}
In this work, I studied the prospects of inferring the WIMP speed distribution in addition to the WIMP particle-physics parameters for several benchmark models from direct-detection data sets. I created mock data sets for idealized versions of cryogenic and liquid-noble experiments expected to be on line by or near 2015.  I applied Bayesian inference to estimate WIMP parameter values and uncertainties from the mock data sets.

There were three cases I considered.  In the first case, I considered constraints on WIMP particle-physics and speed-distribution parameters in the case that hypothesis for the speed distribution matched the actual form of the speed distribution, but for which the parameters of the speed distribution were otherwise unconstrained.  The motivation for this study was twofold.  First, most parameter forecasts for direct-detection experiments have focused on the hypothesis of a smooth halo WIMP DF, with the speed parameters \vlag~and \vrms~fixed to something like the SHM \cite{green2010,pato2010}.  If the speed-distribution parameters were allowed to vary at all, it was typically not over a wide range.  Thus, I wanted to explore a number of benchmark halo DF scenarios with weak priors on the parameters to see how well one could infer both the speed-distribution and WIMP particle-physics parameters.  Although Sec. \ref{subsec:1mb} focused on benchmark speed-distribution models that spanned a reasonable range for a smooth halo hypothesis, I have also considered other single-mode models (e.g., if the dark disk or a single large velocity stream dominates the local DF) and found similar results.  Second, the best parameter constraints are obtained when the hypothesis for the speed-distribution model is correct, and I wanted to know how good those best constraints are likely to be for a variety of benchmark speed-distribution and particle-physics parameter sets.

I found that for any benchmark speed-distribution model, that one could get reasonable constraints on \mwimp~and $D$ with only the weakest of priors on the theoretical parameters, and that constraints improved significantly if \Qmax~was set quite high.  Constraints were tightest for \mwimp$\lesssim 100$ GeV and for \vlag$< 280\hbox{ km s}^{-1}$.  In general, the degree of uncertainty in \mwimp~and $D$ depends on the underlying speed distribution.

The constraints on \vlag~and \vrms~depend on the underlying speed distribution as well as \mwimp~and $D$.  Constraints were generally tighter for smaller \mwimp~and \vrms.  In general, \vrms~is far better constrained than \vlag; it is only possible to place an upper limit on \vlag.  This was true even when I introduced a strong prior on the WIMP mass, a prior of the sort one would expect if supersymmetry were discovered at the LHC.  This prior sharpened constraints on \vrms~but not \vlag~unless \mwimp$\gtrsim 100$ GeV.  The somewhat sobering conclusion is that while it is possible to get good constraints on the velocity dispersion of WIMPs, it will be significantly harder to determine the typical speed of WIMPs with respect to the Earth with 2015-era experiments.   

Second, in Sec. \ref{subsec:2mb} I considered what constraints one would obtain under the hypothesis of a Maxwell-Boltzmann distribution in the case that the true DF were multimodal.  The motivation for studying this case is the strong theoretical prior on the form of the WIMP DF that usually goes into direct-detection parameter forecasts.  For both the SHM + SDD and SHM + SDD + 2 streams multimodal DF models, I found that \mwimp~and $D$ were biased low by $\sim 50\%$, mostly due to the dark-disk component, something that was also found by Ref. \cite{green2010}.  Although these offsets are not enormous for these particular WIMP DF models, they could be more severe for other models.  Moreover, we lose information about the nature of the WIMP speed distribution, such as its multimodal character, if we restrict ourselves to a single-velocity-component hypothesis.

Finally, in Sec. \ref{subsec:vel} I considered a simple empirical model for the speed distribution, both to get a sense of how well one may recover WIMP particle-physics and speed-distribution properties as well as its use for Bayesian model selection.  I showed that the five-bin step-function speed-distribution model successfully reproduced the WIMP speed distribution for speeds above $v = 200 \hbox{ km s}^{-1}$, and that the bias in the $v=0-200\hbox{ km s}^{-1}$ bin was due to the experiments' lack of sensitivity to WIMPs with $v \lesssim 100\hbox{ km s}^{-1}$.  For all benchmark speed-distribution models I considered, the inferred values of \mwimp~and $D$ were biased relative to their true values, but no more so than with the single Maxwell-Boltzmann hypothesis used in Sec. \ref{subsec:2mb}.  However, the biases were less severe if the analysis window was extended to higher energies.  Moreover, there were hints from the ten-bin step-function model that the bias was less than for the five-bin model, even though the Occam's razor philosophy of the Bayesian model-selection criteria I considered favored the five-bin model.  The bias should decrease as better empirical speed-distribution hypotheses are found.

I found that Bayesian model-selection criteria were largely successful in ranking hypotheses for the speed distribution.  For the SHM benchmarks, the Bayesian model-selection criteria indicated that the Maxwell-Boltzmann hypothesis was a better fit to the data than the step-function hypothesis, although the significance of the preference was not strong.  For the multimodal benchmarks, these model-selection criteria showed moderate to strong preference for the step-function model over the Maxwell-Boltzmann model.  Moreover, they showed that doubling the number of bins in the step-function model did not improve the fit over the five-bin model enough to justify the additional bins, except for the one case mentioned at the end of Sec. \ref{subsec:vel}.  These findings are interesting for several reasons.  First, they show that it is possible to get reasonable constraints on the speed distribution with few-parameter empirical speed-distribution hypotheses.  This is good because we do not \emph{really} know what to expect for the WIMP speed distribution.  Second, they show that it is possible to rule in or out popular theoretical models for the WIMP DF.

\subsection{Outstanding issues}
Although I have shown that it is possible to distinguish between Maxwell-Boltzmann and step-function WIMP speed-distribution hypotheses with a modest amount of direct-detection data, there are still a number of questions regarding model selection for the speed distribution.

First, although I have shown that the five-bin step-function speed-distribution hypothesis yields reasonable constraints on the speed distribution and only moderately biased constraints on \mwimp~and $D$, I have not shown that it is the best hypothesis for the speed distribution.  In fact, it cannot be the best hypothesis, since the best hypothesis would be that which matched the form of the benchmark speed distributions.  Moreover, its discontinuous nature is both nonphysical and creates strange features in the posterior, such as the multimodality in the $m_\chi-D$ constraints in Figs. \ref{fig:vel_msigma} to \ref{fig:vel_4vel_msigma}.  However, since we do not \emph{really} know what to expect for the WIMP speed distribution, it is best to explore a variety of empirical models.  In this work, I considered equal-sized (in speed) step functions to model the speed distribution, but it is possible that allowing the widths of the step function to vary  or choosing smoother, continuous basis functions is better.  For example, in Sec. \ref{subsec:vel} I showed that $g(v)$ below $v\sim 100\hbox{ km s}^{-1}$ is not well constrained due to the thresholds of the experiments, but that bins of width $100 \hbox{ km s}^{-1}$ were too small.  Perhaps a better strategy would be to have a single bin for speeds for which the typical recoil energy lies below threshold, and other-sized bins for higher speeds, or to expand the speed distribution into a set of orthogonal functions.  Reference \cite{drees2007} uses overlapping step-function bins of various sizes.  The question is, what is the best strategy to search through these possibilities?  This will depend on the true values of the WIMP particle-physics parameters and the speed distribution, but it is worth putting some thought into how to find empirical hypotheses that will maximize our return on the data.

Finding a good hypothesis for the speed distribution is important not just for the sake of constraining the speed distribution, but a better hypothesis for the speed distribution should also lead to less biased inferences for the particle-physics properties of WIMPs.  As demonstrated in Sec. \ref{subsec:vel}, even though the step-function hypothesis leads to good constraints on the speed distribution, the inferred values of the particle-physics parameters are offset from their true values.  I will delve into this topic more in Sec. \ref{subsec:lhc}.   

However, there are also several theoretical models for the speed distribution that would be interesting to test using direct-detection data.  So far, I have only considered Maxwell-Boltzmann theoretical models, which are based on arguments along the lines of those found in Sec. \ref{subsec:astro} and Appendix A of Ref. \cite{peter2008}.  There are other theoretical models for the local speed distribution that are based on N-body simulations\cite{lisanti2011}, in particular, for the ansatz that the speed distribution is dominated by a smooth halo component.  It would be interesting to use the best empirical models for hypothesis testing against a wider class of specific theoretical models.  For example, if none of the smooth halo predictions fit better than the best empirical fit, then this might suggest that the WIMP speed distribution is multimodal or that the velocity distribution is anisotropic.  Multimodal speed distributions are a signature of the Milky Way's accretion history, and it would be interesting to determine how much one could learn about the accretion history based on the direct-detection data.

An interesting question is if one may infer the escape speed of WIMPs from the Milky Way.  Currently, the best constraints on the local escape speed come from measurements of the radial velocities of local high-velocity stars with the RAVE survey \cite{smith2007}.  The 90\% confidence limits for the escape speed are $498\hbox{ km s}^{-1} < v_{\mathrm{esc}} < 608 \hbox{ km s}^{-1}$ and are somewhat model dependent.  While there is likely a population of WIMPs passing through the Solar System that are unbound to the Galaxy due to the fact that the Galaxy is still accreting matter, it is probably small since the Sun sits deep inside the halo.  If there is a sharp drop off in the distribution function at the escape speed, this should leave an imprint in the nuclear recoil spectra.  The question is if this imprint is large enough to discern even for optimistic WIMP particle-physics and speed-distribution scenarios.  A further complication is that the types of experiments I considered in this work are not sensitive to direction, so there will be some uncertainty in the mapping of the geocentric speeds to a Galactocentric reference frame.  However, this is an interesting question but I defer a study thereof to future work.

There are a number of more ``practical'' issues I have not addressed yet.  First, I have ignored backgrounds, energy uncertainties, and systematics.  Obviously, these experimental realities will affect parameter estimation.  Furthermore, I have also ignored several theoretical issues beyond the ansatz of a WIMP model for dark matter.  For example, I made the ansatz that \sigmapsd$=$\sigmansd$=0$, which is almost certainly not the case in reality.  There are also uncertainties on the form factor $F_{SI}$.  The uncertainties are even greater for spin-dependent scattering \cite{jungman1996,ellis2008}.  One way forward is to parametrize all the uncertainties, backgrounds, and systematics; throw them all into a likelihood function; and search a greatly expanded parameter space with the use of the types of Bayesian tools I used in this work.

\subsection{Complementarity with other data sets}\label{subsec:lhc}
Beyond finding a good model for the speed distribution for its own sake, it is useful to characterize the speed distribution well for the purpose of WIMP identification.  Once WIMPs are discovered through multiple channels (e.g., produced at the Large Hadron Collider, inferred from the observed shower of particles from WIMP annihilation), one will want to see if the particles discovered through these channels are actually of the same type.  In addition, if the same WIMP particle is responsible for all these signals, one will want to characterize the WIMP and the theory to which it belongs using the data.  There have been several studies to investigate how data from the LHC and direct-detection experiments can be used to constrain specific theories for physics beyond the standard model, especially supersymmetry \cite{baltz2006,altunkaynak2008,bertone2010,goodman2010}.  Given the high complexity of the parameter spaces of these theories, a wide network of points in the parameter space can yield the same set of LHC observables.  This can lead to estimates of the relic density and spin-independent cross sections that span orders of magnitude.  It is especially important to be able to estimate the relic density for a theory given the data because the one thing we know about dark matter to a high degree of accuracy is the its abundance in the Universe (e.g., \cite{breid2010,komatsu2011}).  If a theory that fits the data reasonably well results in a too-large relic abundance of dark matter, then it is ruled out.  If the theory predicts a relic abundance that is significantly below the true relic abundance, it indicates that the dark matter created in the collider is only a subdominant component of dark matter as a whole.

References \cite{baltz2006} and \cite{bertone2010} have shown that the addition of direct-detection data can significantly improve constraints on the estimated relic abundance given the collider data and a specific theory.  However, these authors fixed the WIMP distribution function in their analyses.  As I have shown in Sec. \ref{sec:results}, a poor hypothesis for the speed distribution will result in biases in \mwimp~and $D$.  Even in Sec. \ref{subsec:vel} in which the step-function speed-distribution hypothesis was a significantly better fit to the multimodal distribution-function data than the Maxwell-Boltzmann hypothesis, it still lead to biases in the WIMP particle-physics parameters.  If these speed-distribution-dependent biases are not fully understood, they could lead to incorrect inferences about the WIMP particle model.

In order to facilitate accurate joint analyses among collider, direct-detection, and indirect-detection data sets, I recommend performing parameter inference for a variety of speed-distribution hypotheses, including several empirical models.  This will give us a sense of how the uncertainty in the speed-distribution model affects, for example, the inferred relic density for a specific particle model.

\section{Conclusion}\label{sec:conclusion}
In this work, I have created and analyzed mock data sets for idealized 2015-era liquid-noble and cryogenic direct-detection experiments.  The main point of this work was to explore how well one might reconstruct the WIMP speed distribution in addition to the particle-physics properties of WIMPs (mass, cross sections) from future data sets using Bayesian inference.  The main findings are the following:
\begin{itemize}
  \item Regardless of the true WIMP distribution function or of the hypothesis for the form of the WIMP distribution function, it is unnecessary at best and misleading at worst to place strong priors on the parameters of the distribution-function model.  Even using extremely weak priors on the WIMP mass, cross sections, and distribution-function parameters, one is able to get good constraints on all parameters for data sets as small as 200 events for all experiments combined.  If, however, there are significantly fewer events, parameter constraints will be prior dominated if the prior is strong.
    \item The constraints improve significantly if the analysis windows for the direct-detection experiments are extended as high in energy as possible.
    \item Empirical speed-distribution hypotheses lead to good reconstruction of the WIMP speed distribution.  I recommend further investigation into their use for direct-detection parameter inference.
    \item  Even for the modest mock data sets in this work, it is possible to use Bayesian model-selection criteria determine if a specific functional form for the distribution function fits better than a simple empirical speed-distribution model.  This is especially useful for determining if the local WIMP population is dominated by an equilibrium dark-matter halo population or bears significant imprints of the Milky Way's accretion history. 
\end{itemize}

\begin{acknowledgments}
I thank Vera Gluscevic, Kim Griest, Greg Martinez, Bernard Sadoulet, Pat Scott, and Roberto Trotta for useful discussions, and the Aspen Center for Physics for its hospitality while part of this work was completed.  This work made use of the University of California, Irvine Greenplanet computing cluster.  This research was supported by the Gordon and Betty Moore Foundation, the Center for Cosmology at UC Irvine, NASA Grant No. NNX09AD09G, and NSF Grant No. 0855462.
\end{acknowledgments}


\end{document}